\documentclass[aps,prb,noshowpacs,amsmath,amssymb,twocolumn,longbibliography,superscriptaddress]{revtex4}
\usepackage[utf8]{inputenc}
\usepackage{graphicx} 
\usepackage{amsmath}
\usepackage{amssymb}
\usepackage{xcolor}
\usepackage{color} 
\usepackage[abs]{overpic}
\usepackage{epstopdf}
\usepackage{soul}
\usepackage{mhchem}
\DeclareUnicodeCharacter{2212}{-}
\usepackage{hyperref}

\usepackage{ulem}

\definecolor{redred}{HTML}{D53E4F}

\definecolor{greengreen}{HTML}{2F9B31}

\graphicspath{ {./images/} }

\begin{document}

\title{Boundary time crystals in collective $d$-level systems }
\author{Luis Fernando dos Prazeres}
\affiliation{Instituto de Física, Universidade Federal Fluminense, Niterói, Rio de Janeiro, Brazil}

\author{Leonardo da Silva Souza}
\affiliation{Instituto de Física, Universidade Federal Fluminense, Niterói, Rio de Janeiro, Brazil}

\author{Fernando Iemini}
\affiliation{Instituto de Física, Universidade Federal Fluminense, Niterói, Rio de Janeiro, Brazil}

\date{\today}

\begin{abstract}
Boundary time crystals (BTC's) are non-equilibrium phases of matter 
occurring in quantum systems in contact to an environment, for which a macroscopic fraction of the many body system breaks time translation symmetry. We study BTC's in collective $d$-level systems, focusing in the cases with $d=2$, $3$ and $4$. We find that BTC's appear in different forms for the different cases.
We first consider the model with collective $d=2$-level systems [presented in Ref.\cite{Iemini2018}], whose dynamics is described by a Gorini-Kossakowski-Sudarshan-Lindblad master equation, and perform a throughout analysis of its phase diagram and Jacobian stability for different interacting terms in the coherent Hamiltonian. In particular, using perturbation theory for general (non Hermitian) matrices we obtain analytically how a specific $\mathbb{Z}_2$ symmetry breaking Hamiltonian term destroys the BTC phase in the model. Based on these results we define a $d=4$ model composed of a pair of collective $2$-level systems interacting with each other. We show that this model support richer dynamical phases, ranging from limit-cycles, period-doubling bifurcations and a route to chaotic dynamics. The BTC phase is more robust in this case, not annihilated by the former symmetry breaking Hamiltonian terms. 
The model with collective $d=3$-level systems is defined similarly, as competing pairs of levels, but sharing a common collective level. The dynamics can deviate significantly from the previous cases, supporting phases with the coexistence of multiple limit-cycles, closed orbits and a full degeneracy of zero Lyapunov exponents.
\end{abstract}


\maketitle

\section{Introduction}

The classification of different phases of matter according to their spontaneous symmetry breaking (SSB) is a cornerstone of physics and one of Landau's legacy \cite{Goldenfeld:1992qy,sachdev_2000}. It is based on the idea that the system, in the thermodynamic limit, can break some of its symmetries due to thermal or quantum fluctuations giving rise 
 to different phases of matter, as \textit{e.g.}  crystals in case a spatial translational symmetry is broken, superfluids for gauge symmetries, ferromagnets in the case of a rotational spin invariance, among many other different phases.
Recently the existence of a different case of SSB phase (which has intriguingly not been considered until recent years) breaking the time translational symmetry has been under large discussion.
These phases first addressed by Wilczek in $2012$ \cite{wilkczek2012} (and later termed as time crystals) generated an intense debate \cite{Li2012,bruno2013a,bruno2013b,bruno2013c,philippe2013,Volovik2013} and were soon ruled out in thermal equilibrium system by a no-go theorem \cite{watanable2015} (for short-range interacting system) indicating in this way that the proper ground for its existence are in out-of-equilibrium conditions.
In fact, theoretical studies along this direction have been successful in predicting the existence of time crystals in disparate different systems, ranging 
from closed to open systems breaking a continuous or discrete time translational symmetry \cite{Iemini2018,PhysRevB.95.214307,PhysRevB.99.104303,sacha2015,
sachaprl2017,Prokof2018,
nayak2016,khemani2016,sondhi2016,PhysRevLett.118.030401,PhysRevLett.119.010602,PhysRevLett.120.110603,buca2019,
PhysRevResearch.2.012003,PhysRevLett.123.184301,Yao2020,Hurtado2020,PhysRevA.103.013306,giulia2021,RieraCampeny2020timecrystallinityin,PhysRevResearch.2.022002,Lled__2020,PhysRevA.101.033839,marino2021,Homann2021}.
 In particular, discrete time crystal were observed experimentally in $2017$ in an interacting spin chain of trapped atomic ions \cite{zhang2017} and on dipolar spin impurities in diamond \cite{Choi2017}, soon after their theoretical preditions. Later on other systems were also experimentally observed supporting such peculiar phases of matter \cite{PhysRevLett.120.180603,PhysRevB.97.184301,PhysRevLett.120.180602,PhysRevLett.121.185301}. See Refs.\cite{Sacha_2017,ElseReview2020} for interesting reviews on the topics.

\begin{figure*}
\includegraphics[scale=0.99]{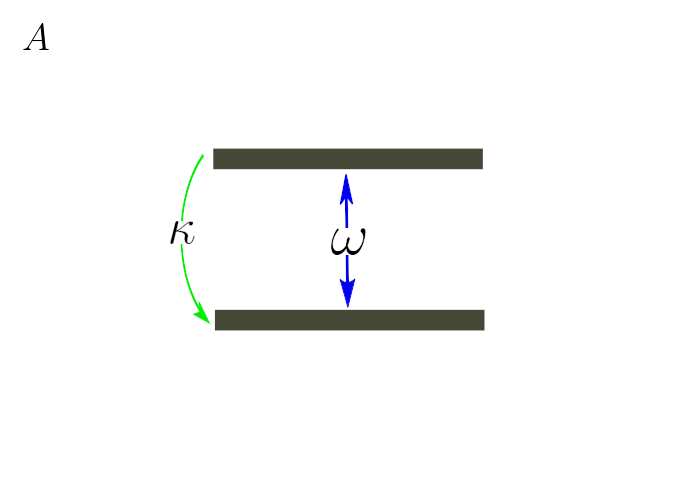}
\includegraphics[scale=0.99]{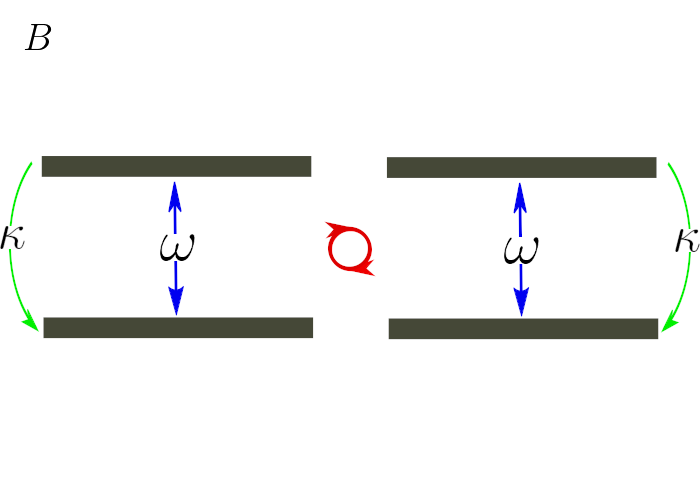}
\includegraphics[scale=0.99]{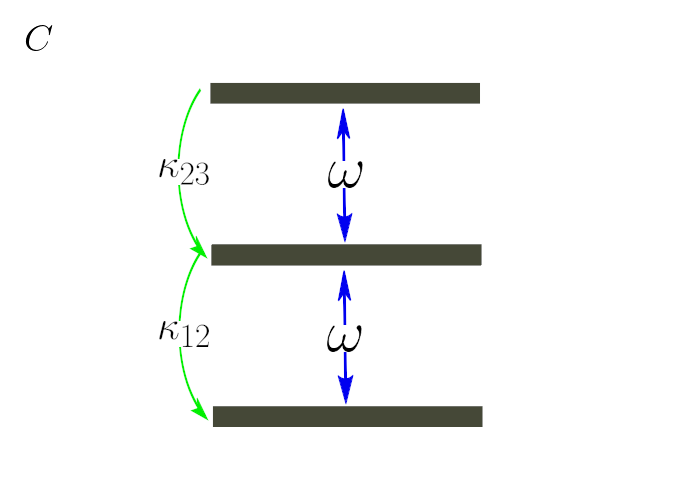}
\includegraphics[width = 0.36 \linewidth]{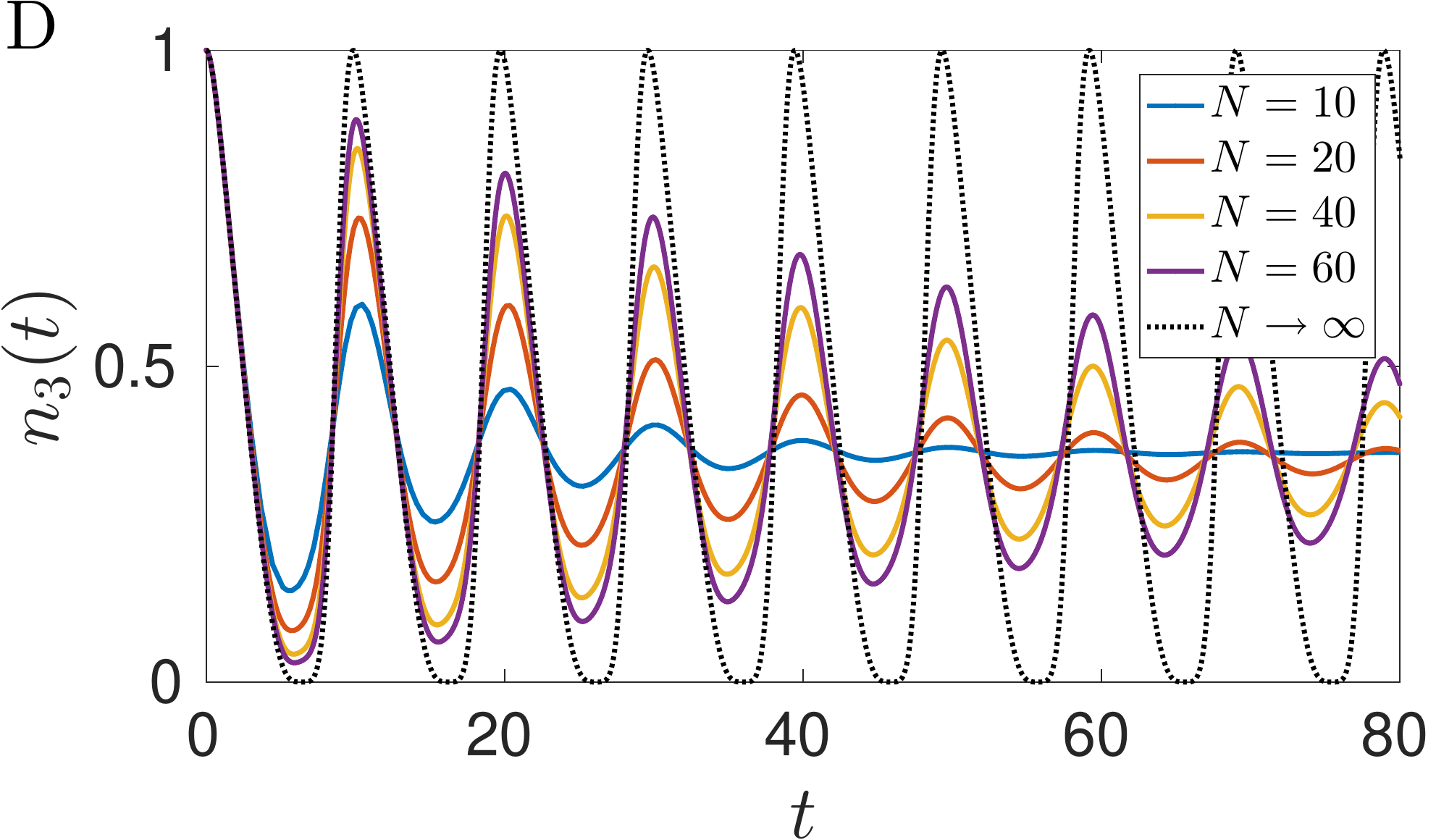}
\includegraphics[width = 0.36 \linewidth]{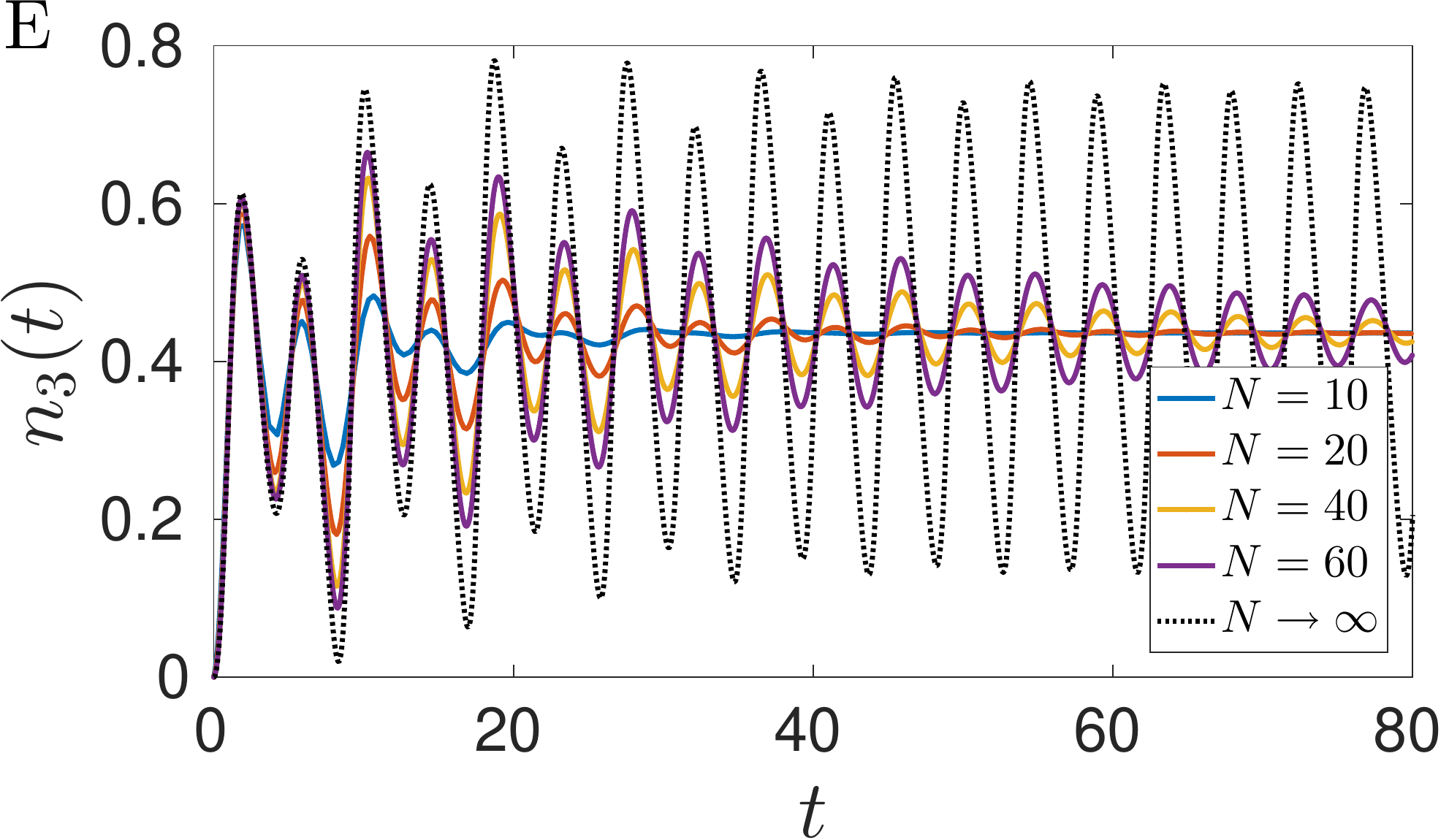}
\includegraphics[scale = 0.31]{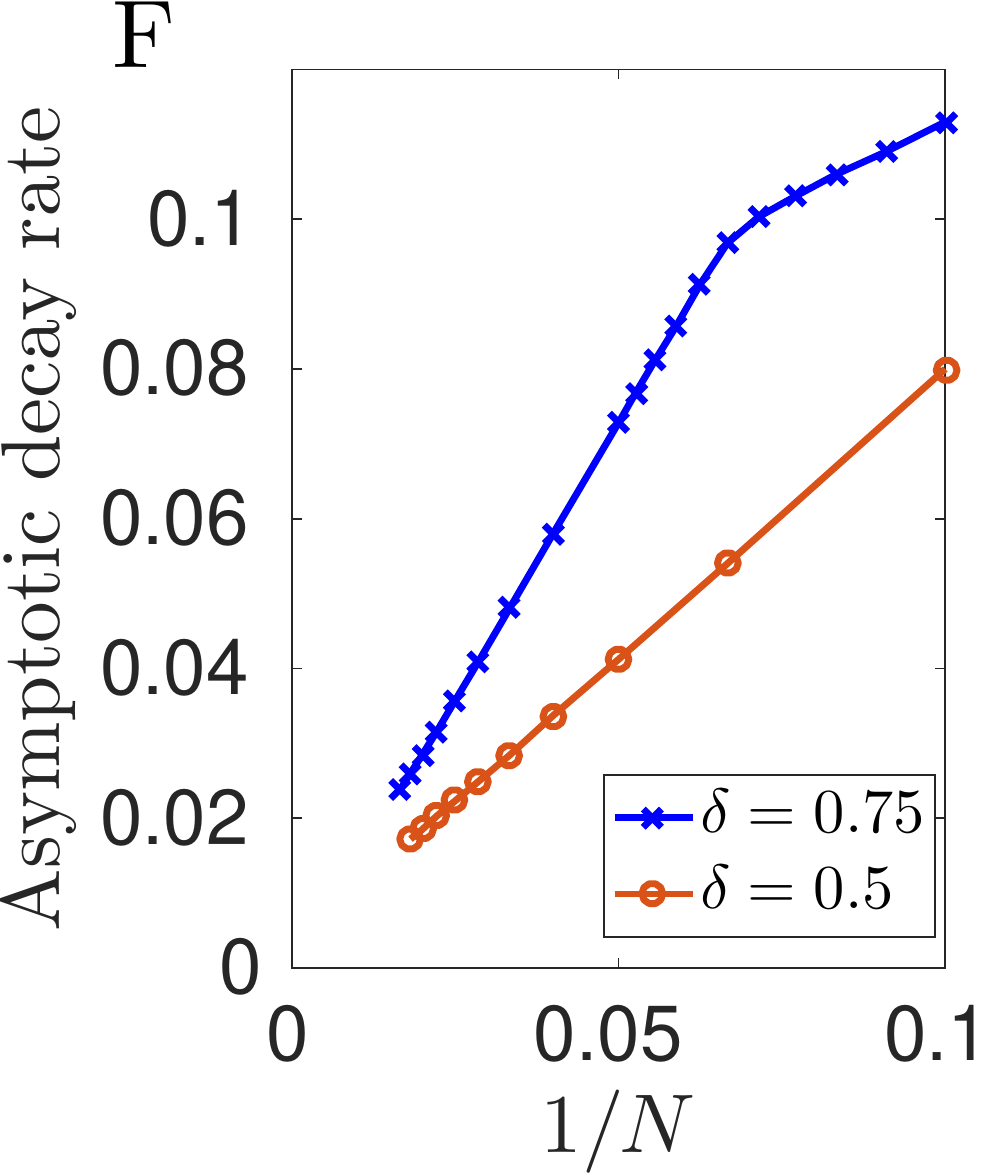}
 \caption{We consider collective $d$-level systems described by a GKS-Lindblad master equation - Eq.\eqref{eq.master.equation} -  
 whose dynamics is driven by coherent Hamiltonian terms ($\omega$) and collective dissipation ($\kappa$). Specifically we study the cases of (A) a collective $2$-level system 
 (Eqs.\eqref{eq.d2.Ham.Lind.definition.eq1}-\eqref{eq.d2.Ham.Lind.definition.eq2}), 
 (B) a pair of interacting collective $2$-level systems 
 (Eqs.\eqref{eq.d4.Ham.Lind.definition.eq1}-\eqref{eq.d4.Ham.Lind.definition.eq2})
 and (C)  a collective $3$-level system 
 (Eqs.\eqref{eq.d3.Ham.Lind.definition.eq1}-\eqref{eq.d3.Ham.Lind.definition.eq2}). In panels (D) and (E) we show two illustrative cases of BTC's dynamics in the $3$-level collective system, featuring periodic orbits (D) or limit cycles (E). While for finite system sizes the collective occupation of level $n_3(t)$ has a finite decay rate (exponential decay rate for the amplitute of the oscillations) and tends to equilibrate to a constant value in the long time limit, in the thermodynamic limit the decay rate vanishes (thus the lifetime of oscillations diverge) and the system shows persistent dynamics. In panel (F) we show the finite-size scaling of the decay rates, featuring an $1/N$ scaling in both cases, thus vanishing in the thermodynamic limit - and only in that limit. We used system parameters 
 $\delta=0.5$ in panel (D),
 $\delta=0.75$ in panel (E) and in both cases $\omega/\kappa = 2,\, \alpha = 1$.
 }
 \label{fig.models}
\end{figure*}

A particularly interesting form of time crystal phases can occur in quantum systems in contact to an environment. In this case the system, also termed as boundary system \cite{Iemini2018}, can break the time translation symmetry while the environment remaining time-translationally invariant. The symmetry breaking appears only at the (macroscopic) boundary system, thus forming a so-called \textit{boundary-time crystal (BTC)} (similar to surface critical phenomena). In such phases the system shows, \textit{only in the thermodynamic limit}, a persistent dynamics of a macroscopic observable breaking the time translation symmetry - see Figs.\eqref{fig.models}-(D-E) for illustrative cases.

A specific case of a BTC was shown in Ref.\cite{Iemini2018} in a model usually used to discuss cooperative emission of radiation 
\cite{Walls1978,DRUMMOND1978160,PURI1979200,Walls_1980}. 
The model consists of $N$ driven $2$-level spins collectively coupled to a boson mode field. The dynamics of the collective spins (boundary system) under a Markovian approximation is described by a time-independent GKS-Lindblad master equation with  competing coherent Hamiltonian driving and a collective decay of spins \cite{Schneider2002,Morrison2008}. For sufficiently high coherent driving strength, the magnetization shows an oscillatory dynamics with lifetime diverging with the number of spins in the system, \textit{i.e.,} while for finite system sizes the collective spin magnetization in the long time limit tends to an equilibrium constant value, in the thermodynamic limit (and only in that limit) the magnetization shows persistent and indefinitely in time oscilations.
The model thus breaks a \textit{continuous} time translational symmetry. Moreover, the dynamics of the collective observables in the BTC phase was shown to possess intriguing properties,
 with a constrained dynamics appearing in the form of closed period orbits with reversibility symmetry and peculiar quasi-conserved quantities. 

An interesting open issue remains in this way a more comprehensive understanding of such peculiar dynamics for the collective $2$-level BTC, with respect to the role of its underlying dynamical symmetries, its stability due to different perturbations and the generality of temporal dynamics in such collective models. 
We address these issues in this work, in particular we show that in such colletive $2$-level systems BTC's always appear as constrained 
periodic closed orbits for general interactions in the system. Moreover, we show  analytically how the $\mathbb{Z}_2$ symmetry in the coherent Hamiltonian term of the Lindbladian dynamics is crucial in order to stabilize the closed period orbits.  We further extend our studies to more general collective models, those composed of collective $d$-level systems, focusing in the cases with $d=3$ and $4$. Each of the considered models have their own merits, as we discuss in more detail below. While the collective $3$-level system has an specific \textit{nontrivial} collective algebra which cannot be reduced to those of usual collective angular momentum, the model with collective $4$-level systems describes the effects of interactions \textit{between} BTC's, subject also very recently 
investigated experimentally \cite{Autti2020}.
 We show that on these more general models, BTC's do exist and appear with different properties than the simpler collective $2$-level case. In particular, in these cases BTC's can feature not only constrained closed periodic orbits, but also richer dynamics with limit-cycles, period doublings and a route to chaos.
 The study of these different forms of BTC's in quantum systems, with its different facets and peculiar properties could help us to improve our understanding for such non-equilibrium phases of matter, sheding some light on their basic mechanisms (and limitations) and the dependence of their collective algebras and underlying symmetries~\cite{giulia2021}. 
We hopefully expect the results of this work be useful on future developments in the field, unraveling new interesting connections to apparently different concepts \cite{Hurtado2020,Estarellaseaay8892}, the nature of elementary excitations~\cite{PhysRevLett.126.020602} in time-crystal phases, or opening paths towards novel applications of TC's in different fields, as recently proposed for the simulation of quantum complex networks~\cite{Estarellaseaay8892} or in protocols for non-Abelian braidings of Majorana edge modes in quantum computation~\cite{PhysRevLett.120.230405}.

In this work we thus consider different models supporting BTC's in collective $d$-level systems. We first discuss in detail the collective $2$-level system and further use it as a basis for the definition of the  cases, featuring even richer dynamical phases. We  discuss the steady state properties of the collective $2$-level system with a combination of analytical and numerical methods, characterizing their stability from a Jacobian perspective. This allows us to make a correspondence of BTC's to the existence of \textit{centers} (also called neutrally stable fixed points) in the model. From this perspective we trace the phase diagram of the model considering different interacting terms in the Hamiltonian. We also highlight the role of the certain symmetries in the model.

In a second part of the manuscript we use these results as a basis for the definition of an extended model composed of a pair of collective interacting $2$-level systems - Fig.\eqref{fig.models}-B. 
We show that this model supports richer dynamical phases. 
In particular we show that for certain interactions breaking the coherent Hamiltonian symmetry (as well as apparently any quasi-conserved quantity) the model shows limit-cycles regimes, thus supporting a more robust BTC phase.
Further varying the interaction the system tends to a chaotic dynamics from subsequent period-doubling bifurcations. We also analyse the period-doubling bifurcation ratio of the model and its Lyapunov spectrum.

In a third part of the manuscript we consider a model with microscopic constituents composed of $3$-level subsystems (Fig.\eqref{fig.models}-C)  rather than the $2$-levels or pairs of them. The model deviates significantly from the simpler $2$-level case. The collective operators now belongs to an $SU(3)$ algebra, which cannot be reduced to $SU(2)$ or products of it, which has basic implications to the global and dynamical symmetries of the system. We study in particular a Lindbladian with the competition of pairs of $2$-levels, similar to the previous case, however considering now the case in which they share a common level.
 The phase diagram of the model shows static steady states characterized by a ``dark level'', limit-cycle dynamics and a peculiar dynamical phase at a critical line supporting the coexistence of multiple limit-cycles and closed orbits.

This manuscript is organized as follows.
 In Sec.\eqref{sec.models} we define the  three different models studied in this work.
 We start our analysis with the simplest system, \textit{i.e.,} with $d=2$.
 In Sec.\eqref{d2.dynamical.equations} we derive its dynamical equations of motion from a semiclassical approach and discuss its symmetries and quasi-conserved quantities. 
We also obtain analytically the different steady states of the model and analyze 
 in Sec.\eqref{sec.stability.analysis} their linear stability from the Jacobian matrix.
 In Sec.\eqref{sec.symm.breaking.pert}, based on our previous results, we study the effects of a specific symmetry breaking Hamiltonian perturbation in the model and the stability of the BTC's.
 In Sec.\eqref{sec.d4} we start the study of the pair of collective systems, \textit{i.e.,} $d=4$. We first obtain the dynamical equations of motion and discuss their symmetries. 
 In Sec.\ref{sec.route.to.chaos} we explore the effects of interactions between the pair of collective systems, showing the appearance of  limit-cycle regimes, period-doubling bifurcations and a route to chaos.
  In Sec.\eqref{d3.dynamical.equations} we move our analysis to the model with $d=3$. We first introduce the basics of $SU(3)$ algebra and Gell-Mann basis, derive the semiclassical dynamical equations and discuss its symmetries. In Sec.\eqref{d3.dynamical.equations} we analyse the phase diagram of model. We present our conclusions in Sec. \eqref{sec.conclusion}.

\section{The models}
\label{sec.models}

In this section we define the models studied in the manuscript and its general properties.
Inspired on the simplest $2$-level model,  we consider extended systems composed of $d$-level subsystems collectively coupled to a common Markovian environment, leading to a time-independent Lindbladian master equation evolution \cite{nielsen2000quantum},
\begin{equation}\label{eq.master.equation}
    \frac{d}{dt} \hat\rho = \mathcal{\hat L}[\hat \rho] =  i [\hat\rho ,\hat H] + \sum_i \left(\hat L_i \hat\rho  \hat L_i^\dagger + \frac{1}{2}\{\hat L_i^\dagger \hat L_i,\hat\rho \}\right),
\end{equation}
with $\mathcal{\hat L}$ the Lindbladian superoperator, $\hat H$ the coherent driving Hamiltonian of the system and 
$\hat L_i$ the Lindblad jump operators, describing the coupling of the system to the environment.
We are mainly focused in the analysis of  differents forms BTC’s appearing on collective systems within a \textit{theoretical bias}, in order to unveil the generality of temporal dynamics in such Lindbladian systems, the dependence on their collective algebra and underlying symmetries. Therefore we do not perform a through discussion of experimenal implementation of the Lindbladian models. We remark, however, 
that it is always possible to find a Hamiltonian whose dynamics is described by the given GKS-Lindblad equation~\cite{cheboratev1997,gregoratti2001,gough2015,prior2010,Rosenbach2016,SIiemini2018}.
Moreover, possible prospects for an implementation could be envisioned with state-of-art quantum simulation platforms, as
 trapped ions \cite{zhang2017}, artificial qubits in superconducting circuits \cite{Puri2017}, Rydberg atoms \cite{Henriet2020} and color defects in diamond \cite{Angerer2018} where all-to-all interactions have been recently implemented. 
 
\subsection{Collective $d=2$-level systems}
 
We start considering the simpler case with $d=2$. In this case the coherent Hamiltonian and Lindblad jump operators are defined as,
  \begin{eqnarray}\label{eq.d2.Ham.Lind.definition.eq1}
  \hat H &=& \omega_0 \hat S^x + \frac{\omega_x}{S} (\hat S^x)^2 + \frac{\omega_z}{S} (\hat S^z)^2, \\
  \hat L &=& \sqrt{\frac{\kappa}{S}} \hat S_-, \label{eq.d2.Ham.Lind.definition.eq2}
 \end{eqnarray}
where $S=N/2$ is the total spin of the system, $\hat S^\alpha =  \sum_j \hat \sigma_j^\alpha/2$ with $\alpha = x,y,z$ are collective spin operators, $\hat S_\pm = \hat S^x \pm i S^y$ and 
$\hat \sigma_j^\alpha$ are the Pauli spin operators for the $j$'th subsystem.
The collective operators inherit the $SU(2)$ algebra of their components, satisfying in this way the commutation relations $[\hat S^\alpha,\hat S^\beta] = i\epsilon^{\alpha \beta \gamma S^\gamma}\hat S^\gamma$.
 Due to the collective nature of the interactions, the model conserves the total spin $S^2 = (\hat S^x)^2 + 
 (\hat S^y)^2 + (\hat S^z)^2$.

The model on its simplest form, with $\omega_x = \omega_z=0$, is commonly used to describe cooperative emission in cavities \cite{PhysRevA.98.042113,Walls1978,DRUMMOND1978160,PURI1979200,Walls_1980} and was recently shown to support a time crystal phase with the spontaneously breaking of time-translational symmetry \cite{Iemini2018}. While in the strong dissipative case $\kappa/\omega_0> 1$ the spins in the steady state tend to align down in the $z$-direction, 
in the weak dissipative case $\kappa/\omega_0 < 1$ the dynamics is characterized by persistent temporal oscillations of macroscopic observables.

\subsection{Collective $d=4$-level systems}

In this case we consider a model describing a pair of collective $2$-level (spin $1/2$) systems. Specifically, we define the coherent Hamiltonian and Lindblad jump operators as follows,
 \begin{eqnarray}\label{eq.d4.Ham.Lind.definition.eq1}
   \hat H &=& \frac{\omega_{xx}}{S} \hat S^x_1 \hat S^x_2 + 
 \frac{\omega_{zz}}{S} \hat S^z_1 \hat S^z_2 +
 \sum_{p=1}^2 \omega_{x,p} \hat S^x_p + 
 \omega_{z,p} \hat S^z_p, \\
 \hat L_1 &=& \sqrt{\frac{\kappa_1}{S}}\,  \hat S_{-,1},\qquad 
 \hat L_2 = \sqrt{\frac{\kappa_2}{S}} \,\hat S_{-,2}, \label{eq.d4.Ham.Lind.definition.eq2}
 \end{eqnarray}
where $S=N/2$ is the total spin of each collective system, $\hat S^\alpha_p =  \sum_j \hat \sigma_{j,p}^\alpha/2$ with $p=1,2$, $\alpha = x,y,z$ are the collective
spin operators for the $p$'th collective $1/2$-spin system. The operators $\hat \sigma_{j,p}^\alpha$ are the usual Pauli spin operators for the $j$'th spin in the $p$'th collective system, and the excitation and decay operators are defined analogously $\hat S_{\pm,p} = \hat S^x_p \pm i S^y_p$.
The collective operators inherit the $SU(2)$ algebra for fixed $p$, while commuting otherwise: $[\hat S^\alpha_p,\hat S^\beta_{p'}] = i \delta_{p,p'}\epsilon^{\alpha \beta \gamma S^\gamma}\hat S^\gamma_p$.
 Due to the collective nature of the model, it conserves the total spin for each collective spin system $S_p^2 = (\hat S^x_p)^2 + 
 (\hat S^y_p)^2 + (\hat S^z_p)^2$ for $p=1,2$.

In the case of $\omega_{xx} = \omega_{zz} = 0$ there is no coupling between the two collective systems and the  physics reduces to the simpler $d=2$ case. On the other hand, 
if the coupling between the collective systems in nonzero, as e.g. $\omega_{xx} \neq 0$, one may  expect the strengthen of the persistent oscillations in the time crystal phase, since such couplings can induce local spin excitations on each collective system thus enhancing the effect of coherent collective drivings $\hat S^x_{1,2}$. We will discuss in more detail,  in the next sections, the effects of the different terms in the model leading to a richer phase diagram.

\subsection{Collective $d=3$-level systems}

In this case the model describe cooperative evolution of a collection of three-level subsystems ($d=3$).  We study how a pair of collective $2$-level subsystems compete, or hybridize, when they share a common energy level. 
Specifically, we study the competition of two dissipative channels with the Lindbladian given as follows,
\begin{equation}\label{eq.d3.Ham.Lind.definition.eq1}
\hat{\mathcal{L}} = (1-\delta) \hat{\mathcal{L}}_{12} + \delta \hat{\mathcal{L}}_{23},
\end{equation}
 where $0 \leq \delta \leq 1$ and each $\hat{\mathcal{L}}_{mn}$ acts only in the pair of levels \textit{$m$} and \textit{$n$} - see Fig.\eqref{fig.models}.
  The Lindbladians $\hat{\mathcal{L}}_{mn}$ are defined similarly to the $d=2$ case, with coherent Hamiltonian ($\hat H^{(mn)}$) and Lindblad jump operator ($\hat L^{(mn)}$)  given by,
 \begin{equation}\label{eq.d3.Ham.Lind.definition.eq2}
   \hat H^{(mn)} = \omega_{mn} \hat S^x_{mn},  
 \qquad \hat L_{mn} = \sqrt{\frac{\kappa_{mn}}{S}}\,  \hat S_{-,mn}.
 \end{equation}
 where $S=N/2$ and $\hat S^\alpha_{mn} =  \sum_{j=1}^N \hat \sigma_{j,mn}^\alpha/2$ with $\alpha = x,y,z$ and $m,n=1,2,3$ label the pairs of $(m,n)$ levels. The operators $\hat \sigma_{j,mn}^\alpha$ are the usual Pauli spin operators for the $j$'th subsystem in the pair of $(m,n)$ levels. The  collective excitation and decay operators are defined analogously, $\hat S_{\pm,mn} = \hat S^x_{mn} \pm i S^y_{mn}$.
 
 Similar to the previous $d=4$ case, the model here considers the competition of a pair of collective two-level subsystems. A major contrast comes however from the fact that, due to the shared collective level in the $d=3$ case, the collective operators form an SU($3$) algebra, which cannot be reduced to an SU($2$) as in the $d=2$ case, neither to a pair SU($2$) $\otimes $ SU($2$) as in the $d=4$ case. The dynamics are thus expected to be different from the previous cases, \textit{e.g} one can already notice that the total spin for each pair of two levels is not conserved anymore. The conserved quantities in this case are rather different, given by the two independent Casimir operators of the algebra, a quadratic and cubic operator, respectively. While the quadratic can be seen as a vector norm in the space of group operators, the cubic operator is rather non intuitive. We discuss in more detail these operators in Sec.\eqref{d3.dynamical.equations}.

\section{$d=2$: Dynamical Equations of Motion, Symmetries and Steady States } 
\label{d2.dynamical.equations}

 We study in this section the dynamical equations of motions  for the collective $2$-level system, their symmetries and (quasi)-conserved quantities, as well as obtain the steady states of the model for varying couplings in the Lindbladian.

\textit{Dynamical Equations of Motion.- }
 Although our collective system with a finite number of subsystems $N$ do not have persistent temporal dynamics, in the thermodynamic limit
$N \rightarrow \infty$ (and only in that limit) such symmetry can be broken, thus arising as a many-body phase. In particular, in this limit the dynamics of macroscopic observables composed by a normalized sum of local operators, the ones we will be interested in our current analysis, can be represented in the form of a simpler set of non-linear dynamical equations. This emergent dynamics for the macroscopic observables are obtained through a semiclassical approach, which are exact in the thermodynamic limit for initial states satyisfying clustering conditions \cite{Benatti2018} (as \textit{e.g} simple separable pure states).
In order to derive it we first write the dynamics of a general operator $\hat O$ within the Heisenberg picture, 
\begin{equation}
\frac{ d \langle \hat O \rangle }{dt} = i \langle [\hat H,\hat O] \rangle + \sum_i 
 \langle [\hat L_i^\dagger,\hat O ] \hat L_i
  + \hat L_i^\dagger [\hat O ,\hat L_i] \rangle.
\end{equation}

Considering the collective $2$-level operators $\hat S^\alpha$ and using their SU($2$) commutation relations we find the corresponding dynamical equations,
\begin{align}
\frac{d}{dt} \langle \hat S^x \rangle 
         &=- \frac{\omega_z}{S}\left( \langle \hat S^z\hat S^y \rangle  +  \langle \hat S^y\hat S^z \rangle \right)\nonumber \\
	&\quad+\frac{\kappa}{2S}\left( \langle \hat S^z\hat S^x  \rangle + \langle \hat S^x\hat S^z \rangle  +  \langle \hat S^x \rangle \right), \nonumber \\
\frac{d}{dt} \langle \hat S^y \rangle 
         &=-\omega_0 \langle \hat S^z \rangle  +\frac{\omega_z -\omega_x}{S} \langle \hat S^x\hat S^z +\hat S^z\hat S^x \rangle \nonumber \\
	 &\quad+\frac{\kappa}{2S}\left( \langle \hat S^z\hat S^y  \rangle + \langle \hat S^y\hat S^z \rangle  -  \langle \hat S^y \rangle \right), \\
\frac{d}{dt} \langle \hat S^z \rangle 
        &=\omega_0 \langle \hat S^y \rangle  +\frac{\omega_x}{S}\left( \langle \hat S^y\hat S^x \rangle  +  \langle S^x\hat S^y \rangle \right) \nonumber \\
	&\quad- \frac{\kappa}{S}\left(  \langle \hat (S^y)^2 \rangle  +  \langle (\hat S^x)^2 \rangle  +  \langle \hat S^z \rangle  \right) \nonumber.
\end{align}

We define the macroscopic operators as $\hat m^\alpha =\hat S^\alpha/N$. These operators commute in the thermodynamic limit $[\hat m^\alpha, \hat m^\beta] = i\epsilon^{\alpha \beta \gamma} \hat m^{\gamma}/N$ motivating us to perform a second order cumulant approach (semiclassical approach) for their expectation values 
$\langle \hat m^\alpha \hat m^\beta \rangle \cong \langle \hat m^\alpha \rangle \langle \hat m^\beta \rangle$. In this way in the thermodynamic limit the dynamical equations of motion are closed and given by the following set of nonlinear differential equations:
\begin{eqnarray}
\frac{d}{dt}m^x &=& m^z( -2\omega_zm^y  + \kappa m^x), \nonumber \\
\frac{d}{dt}m^y &=& m^z( 2(\omega_z - \omega_ x)m^x - \omega_0 + \kappa m^y), \\
\frac{d}{dt} m^z &=& \omega_0m^y - \kappa ((m^x)^2 + (m^y)^2) + 2\omega_x m^xm^y.  \nonumber
\label{eq:dynamical.equations.single}
\end{eqnarray}
where we use $m^\alpha \equiv \langle \hat m^\alpha \rangle$ to simplify our notation. 

\textit{Symmetries and (Quasi) Conserved Quantities.- }
 We first see that these dynamical equations  conserves the total spin of the system $\mathcal{N} = (m^x)^2 +(m^y)^2+ (m^z)^2$ - as expected since it should accurately describe the Lindbladian dynamics which has explicitly such a symmetry. Moreover, the dynamical equations also have a quasi-conserved quantity for $\omega_z>\omega_x$ given by,
\begin{align}
 \mathcal{R} &= (-i \kappa +2\omega_z) \log \left(i m^x + m^y - \frac{\omega_0 }{(\kappa -2i\omega_z)} \right) \nonumber \\
&\quad - (-i \kappa - 2\omega_z) \log \left(-i m^x + m^y - \frac{\omega_0 }{(\kappa +2i\omega_z)} \right).
\label{eq.quasi.conserved.su2}
\end{align}
Due to the logarithmic function this quantity is defined up to integer multiples of $2\kappa \pi$, and that is the reason we prefer to define it as a quasi-conserved quantity.

It is worth noticing that the dynamical equations have a reversibility symmetry, given by the following transformation 
\begin{equation}
t \rightarrow -t,\quad m^x \rightarrow m^x, \quad
m^y \rightarrow m^y, \quad
m^z \rightarrow -m^z.
\label{eq.reversibility.sym.su2}
\end{equation}
Furthermore we see a specific structure for the $m^x$ and $m^y$ dynamical equations, where the term $m^z$ can be factored out leading to specific conditions for a steady state.

\textit{Steady States.- } In order to obtain the steady states of the system we must solve the fixed points of the dynamical equations $d m^\alpha/ dt=0$ for $\alpha = x^*,y^*,z^*$, where we denote $(x^*,y^*,z^*)$ the solutions for simplicity of notation. The steady states are given by those \textit{physical }fixed points, i.e., those satisfying the norm condition of fixed total spin $\mathcal{N}$. Noticing the specific structure for the dynamical equations of motion, with the $m^z$ term factoring out, we can consider two different cases for the fixed points: (i) ferromagnetic $m^z \neq 0$ and (ii) paramagnetic $m^z = 0$ fixed points, as shown below:

(i) Ferromagnetic fixed points,
\begin{align}
        x^* &= \frac{2\omega_z\omega_0}{4\omega_z(\omega_z - \omega_x)+ \kappa^2},\nonumber \\
        y^* &= \frac{\kappa\omega_0}{4\omega_z(\omega_z - \omega_x)+ \kappa^2}, \label{eq:fixed.point.mzneq0.single}\\
        z^* &= \pm \sqrt{1 - \omega_0 ^2\frac{4\omega_z
        ^2 + \kappa ^2}{\left(4\omega_z(\omega_z - \omega_x) + \kappa^2\right)^2}}, \nonumber 
\end{align}
which correspond to none or a single pair ferromagnetic steady states, depending on the system couplings.

(ii) Paramagnetic fixed points, given by the algebraic condition,
\begin{equation}
y^* = \frac{\kappa}{\omega_0 + 2\omega_x x^*},
\label{eq:fixed.point.mz0.single}
\end{equation}
In this case the steady states come in pairs, as in the previous case, however we can have $0,1$ or $2$ pairs depending on the system parameters (see Apendix~\eqref{sec.append.d2Paramagnetic} for a detailed discussion).

\section{Stability Phase Diagram}
\label{sec.stability.analysis}

In this section we obtain the phase diagram of the model for different interacting terms in the Hamiltonian, from a stability analysis of its steady states. A simple approach is based on the linearization of dynamical equations of motion around the fixed point, which is effectively described by the Jacobian matrix. Specifically, given the set of dynamical equations of motion
$d m^\alpha /dt = f_{\alpha} (m^x,m^y,m^z)$ 
with $\alpha=x,y,z$ and $f_{\alpha}$  a nonlinear function on the variables, the Jacobian is defined by the matrix $(\hat J)_{\alpha \beta} = \partial f_\alpha/ \partial \beta$.
The spectrum of the Jacobian contains information on the stability of the fixed points.

In case the eigenvalue of the Jacobian matrix have negative (positive) real part the fixed point is an attractor (repeller) and robust to nonlinear terms. In case where the eigenvalues have also an imaginary term the dynamics have spirals towards (away) to the fixed point. Fixed points with nonnull eigenvalue real part are usually called as hyperbolic fixed points. 

Another important case is when the eigenvalue of the Jacobian matrix is purely imaginary, leading to persistent periodic closed orbits around the fixed points, according to the linear approximation. In this case the fixed point is denoted as a \textit{center}.
This type of dynamics with closed periodic orbits (exemplified in Fig.\eqref{fig.models}D) are exactly the one found in Ref.\cite{Iemini2018}  and we argument in this work that 
due to the very contrained dynamics (as discussed in Sec.\eqref{d2.dynamical.equations}) of these collective $2$-level models, this is in fact the only form of persistent dynamics (thus a BTC phase) for general interactions in this model. We thus associate the existence of centers in the dynamical equations to the presence of BTC's in this model.
We corroborate our claim studying analytically and numerically the Jacobian spectrum and system phase-space portraits for different interaction terms in the Lindbladian, as we discuss below.


We start our analysis computing the Jacobian matrix for the fixed points of the model, obtaining that,
\begin{align}
J_{11}&=J_{22}=\kappa z^*,\quad
J_{12} = -2\omega_z z^*,
\nonumber \\
J_{13} &= - 2\omega_ z y^* + \kappa x^*,\quad
J_{21} =  2(\omega_z - \omega_x) z^* ,\\
J_{23}&= 2(\omega_z - \omega_x) x^* - \omega_0 + \kappa y^*,\quad
J_{31} = -2\kappa x^* + 2\omega_x y^*,\nonumber \\
J_{32} &= -2\kappa y^* + 2\omega_x x^* + \omega_0,\quad
J_{33} = 0.\nonumber
\end{align}

The determinant of this matrix is null both for the ferromagnetic and paramagnetic fixed points of the model. This results implies that the Jacobian matrix has always at least one completely null eigenvalue, as usual in dynamical systems with conservation laws \cite{Pikovsky2015}. Moreover, we also obtain explictly the other (two) eigenvalues of the Jacobian (see Appendix \eqref{sec.app.stabilityd2}), showing that while the eigenvalues for the ferromagnetic steady states have always a nonnull real part, in the paramagnetic steady states they are purely real, or imaginary. 
It shows that ferromagnetics phases are not prone to stabilize any persistent dynamics in this specific model, being characterized only by attractors/repulsors whose dynamics in the long time limit tends to equilibriate to a constant.
On the other hand, only paramagnetic phases opens the possibility for such phases, as centers, but this may not be necessarily the case, since depending on system couplings their stability can change from a center to hyperbolic point. 

\begin{figure}
\includegraphics[width = 0.95 \linewidth]{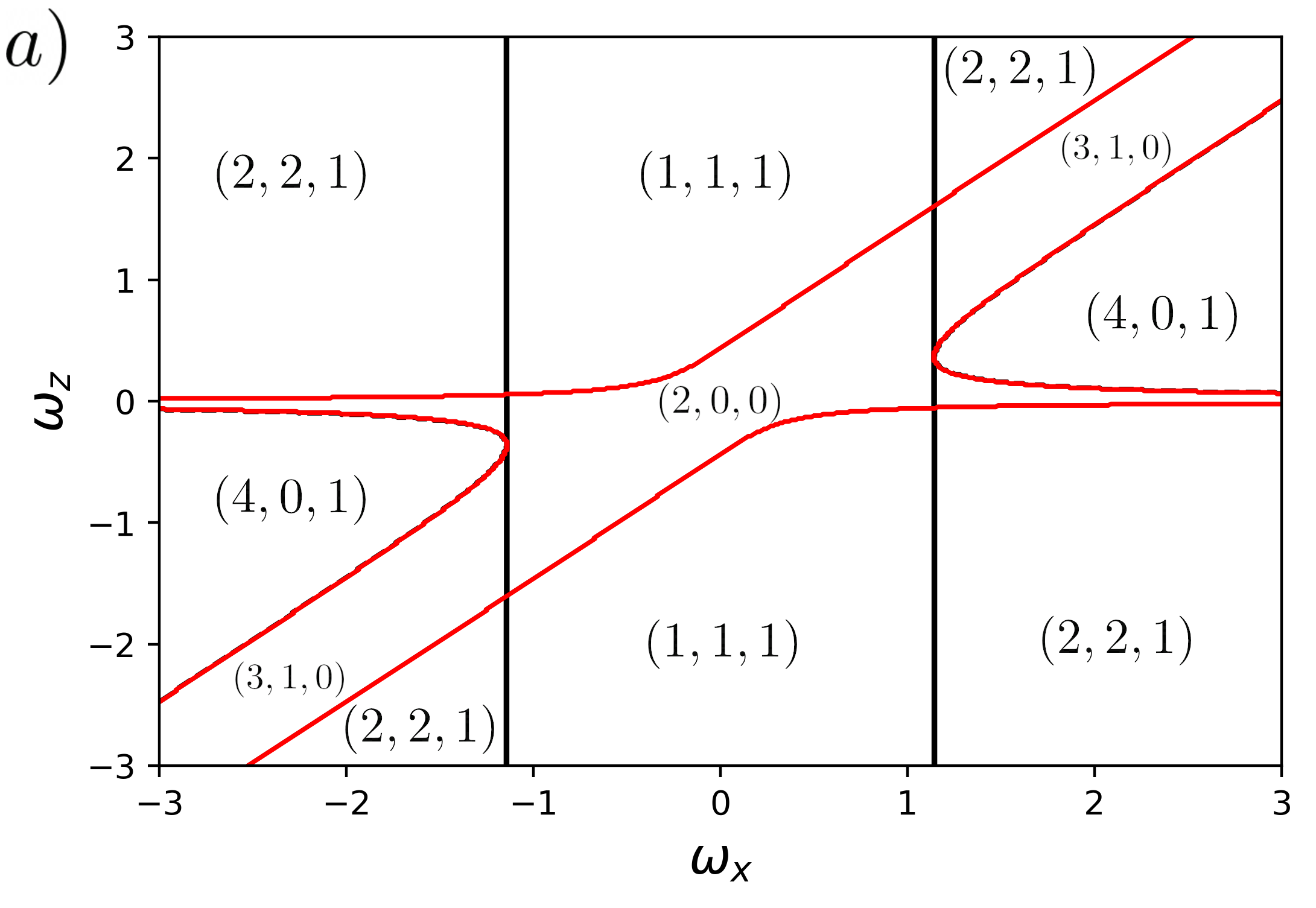}
\includegraphics[width = 0.95 \linewidth]{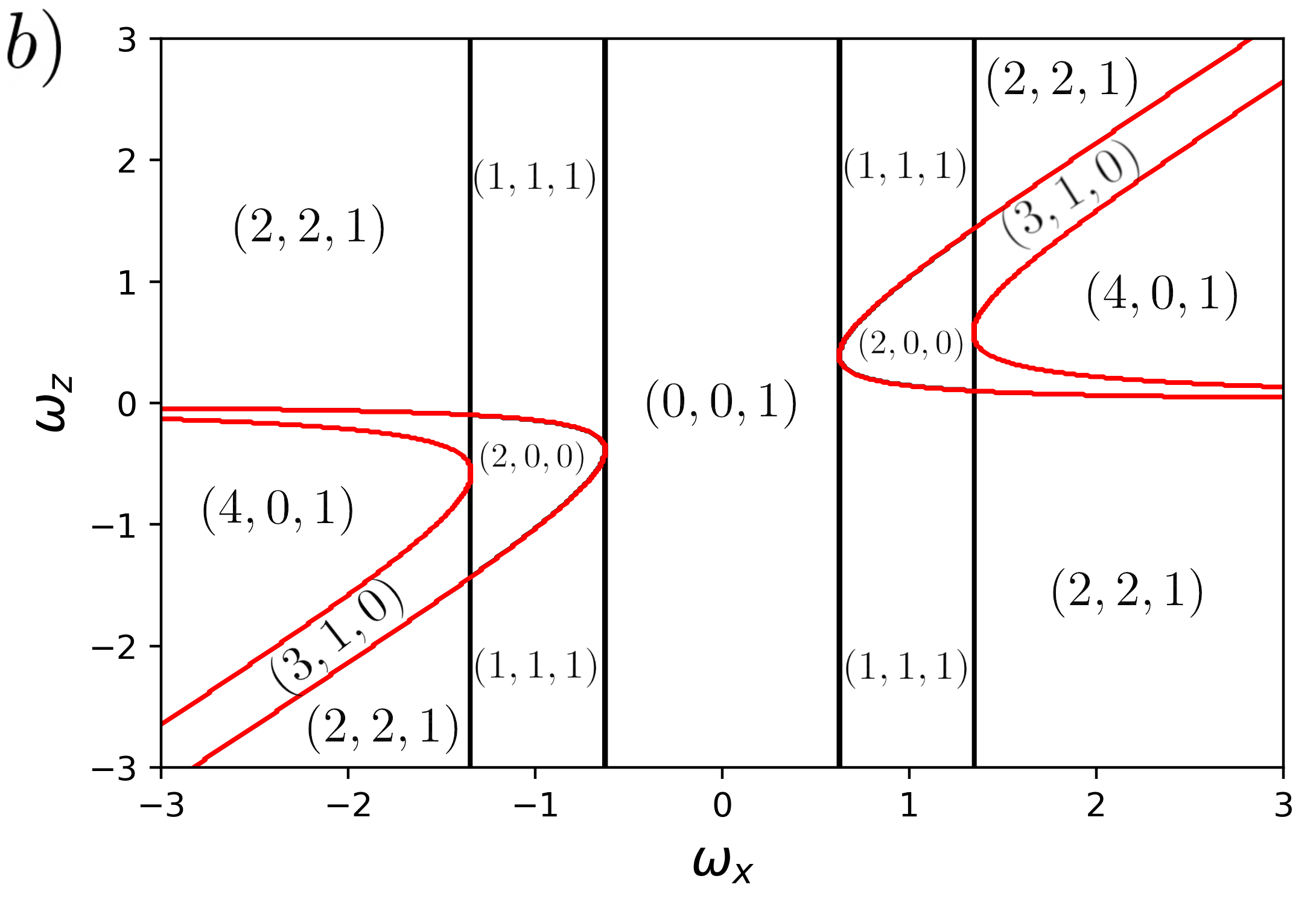}
 \caption{ \textit{Phase diagram from Jacobian analysis:} we show the different phases of the model from the perspective of their steady states stability, for varying Hamiltonian couplings $\omega_z$ and $\omega_z$. We show in \textbf{(a)} our results for a system in the weak dissipative case, with fixed $\kappa = 0.5,\, \omega_0 = 1$ and in \textbf{(b)} for the strong dissipative case, with fixed $\kappa = 1,\, \omega_0 = 0.5$.
   We  describe the phases by the $3$-vector $(n_{\rm pc}, \, n_{\rm ph},\, n_{\rm fh})$, see main text. The red lines highlight when a pair of ferromagnetic steady states are created or annihilated,
$n_{\rm fh} \rightarrow n_{\rm fh} \pm 1$ and simultaneously the stability of paramagnetic steady states are changed,
 $(n_{\rm pc},n_{\rm ph}) \rightarrow  (n_{\rm pc}\pm 1,n_{\rm ph}\mp 1$). The vertical black lines correspond to the creation or annihilation of a pair of  paramagnetic steady states: 
$(n_{\rm pc},n_{\rm ph}) \rightarrow  (n_{\rm pc}\pm 1,n_{\rm ph}\pm 1$)
 }
 \label{fig.phase.diagram.fixed.points}
\end{figure}

\begin{figure*}
\includegraphics[scale=0.22]{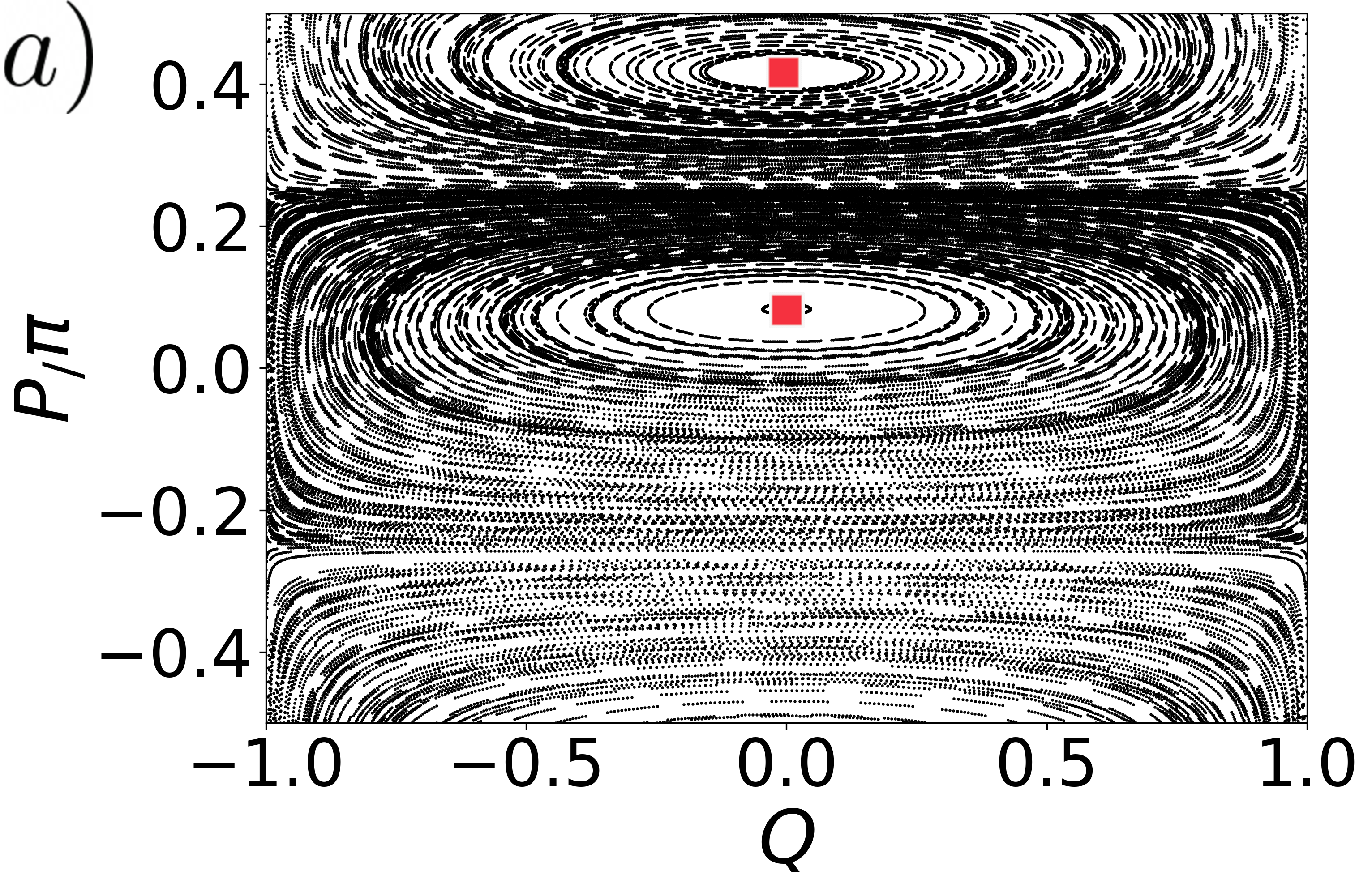} 
\includegraphics[scale=0.22] {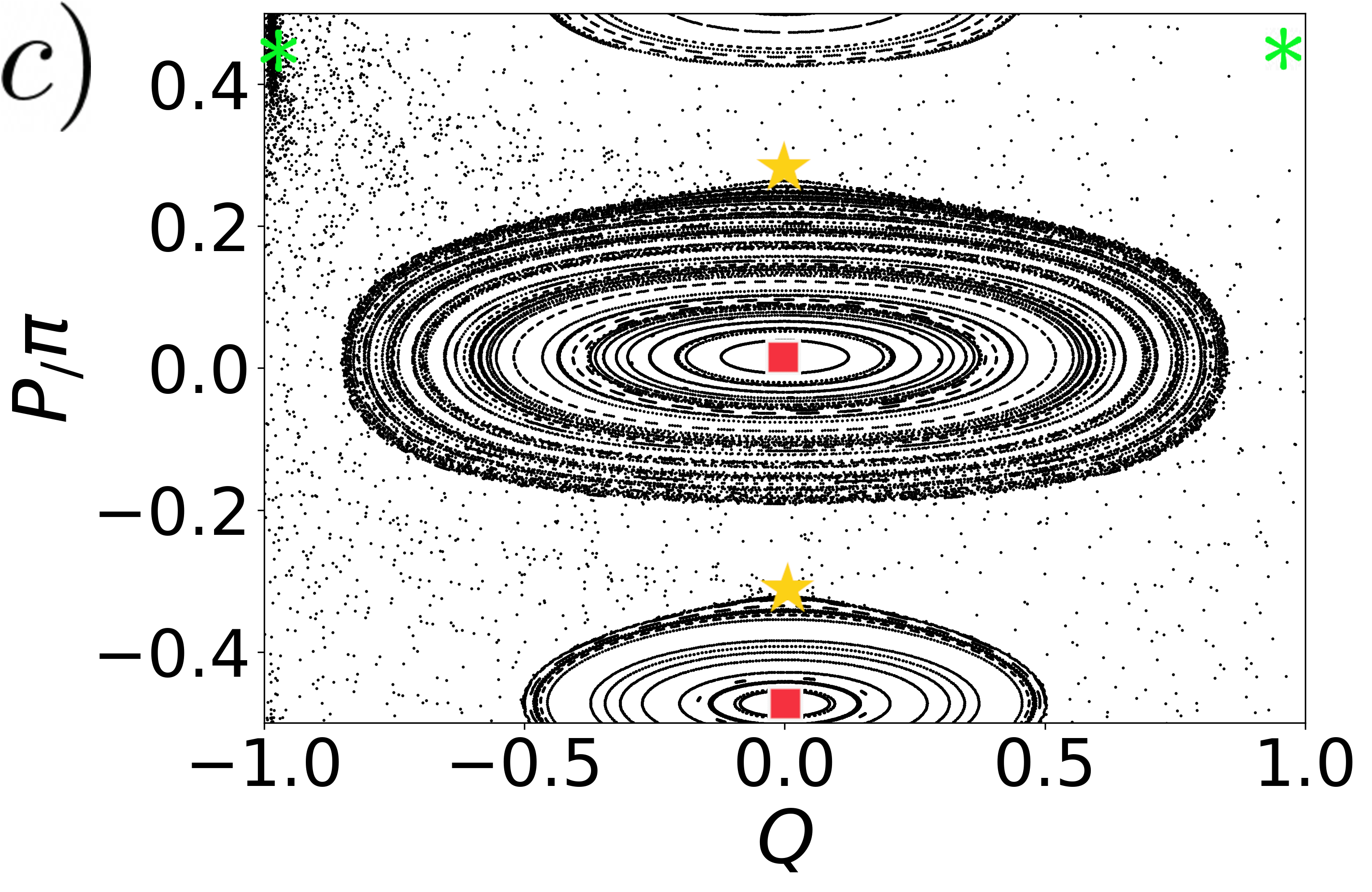}  
\includegraphics[scale=0.22]{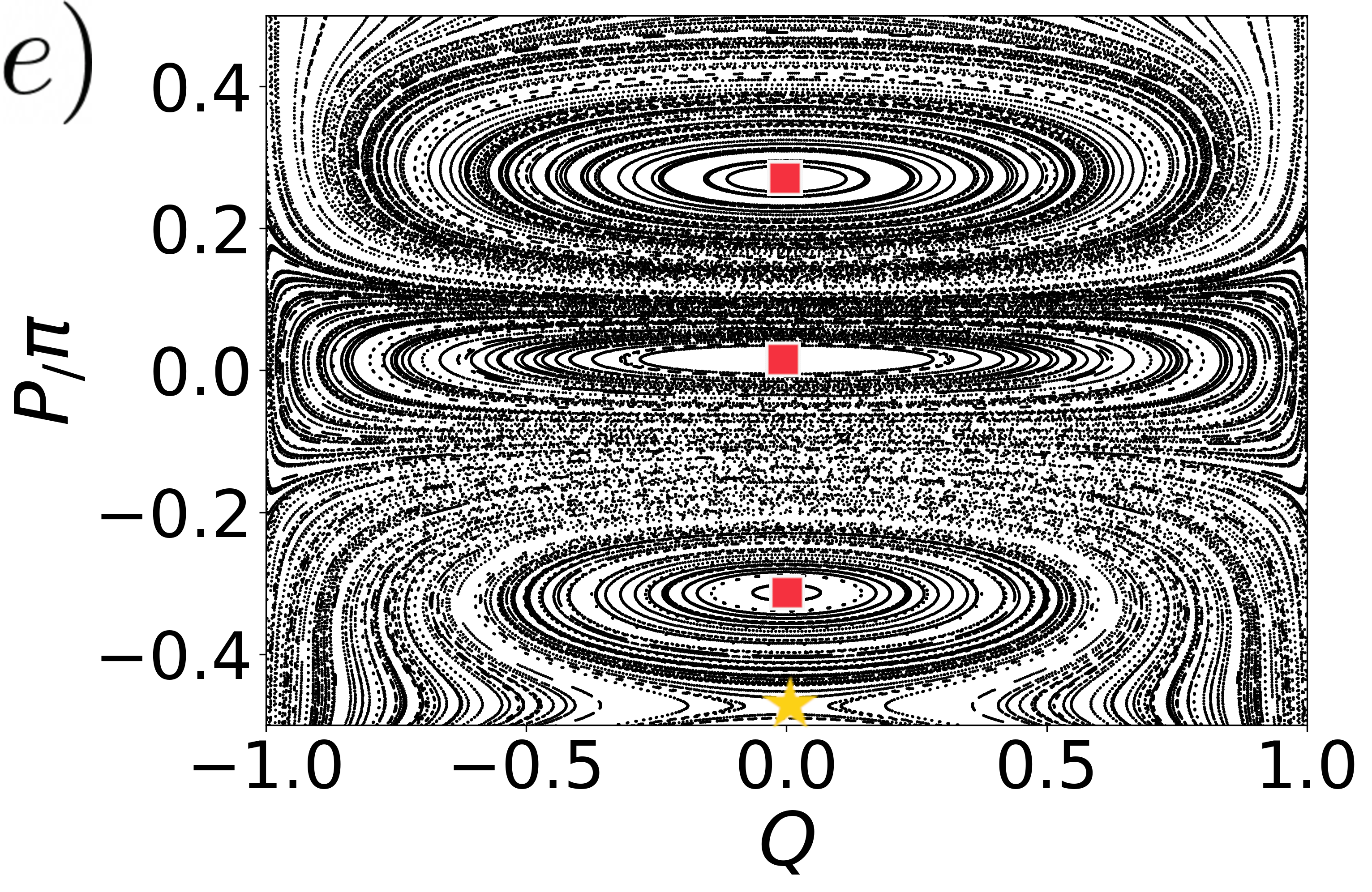} \\
\includegraphics[scale=0.22]{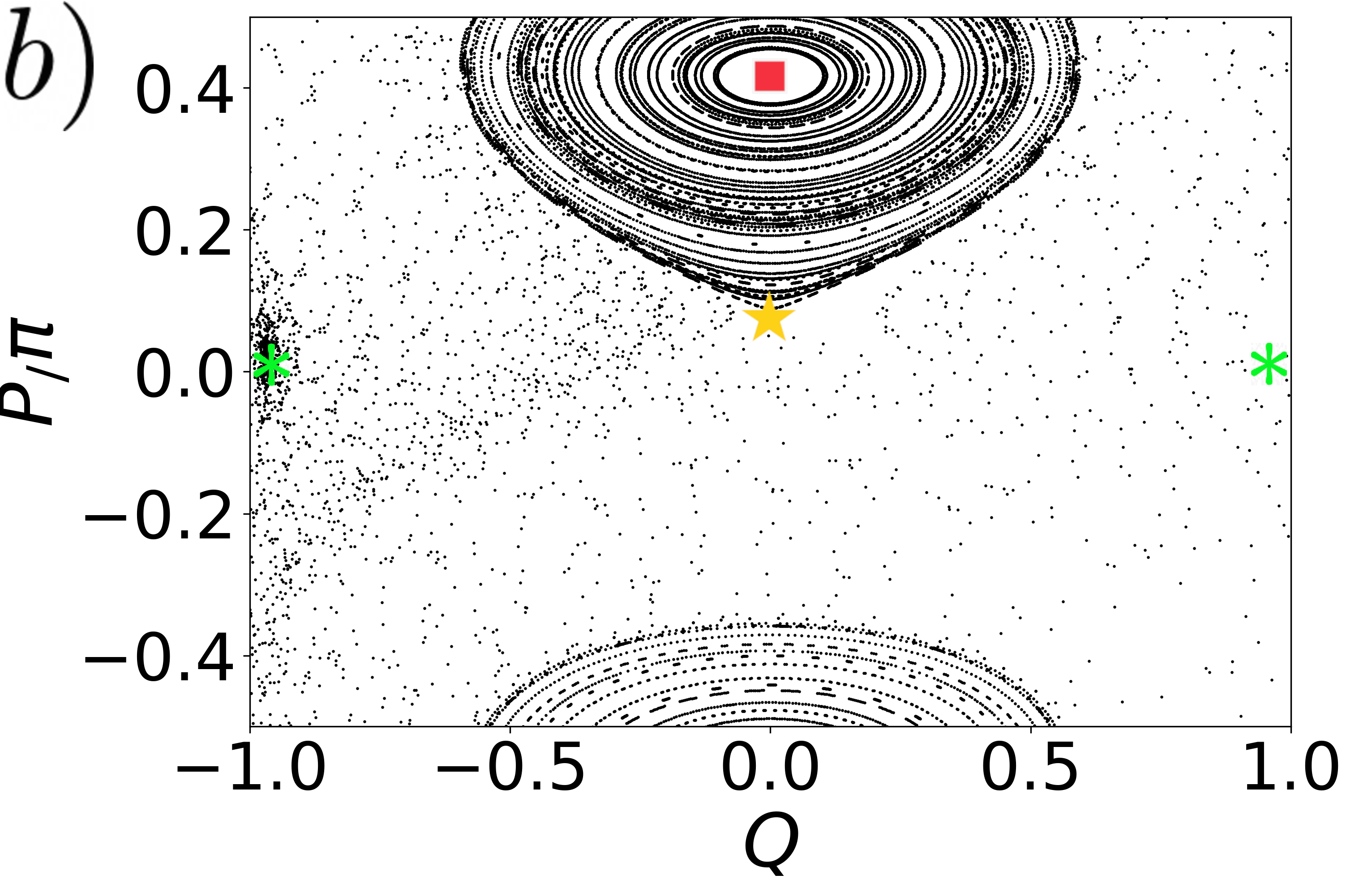}  
\includegraphics[scale=0.22]{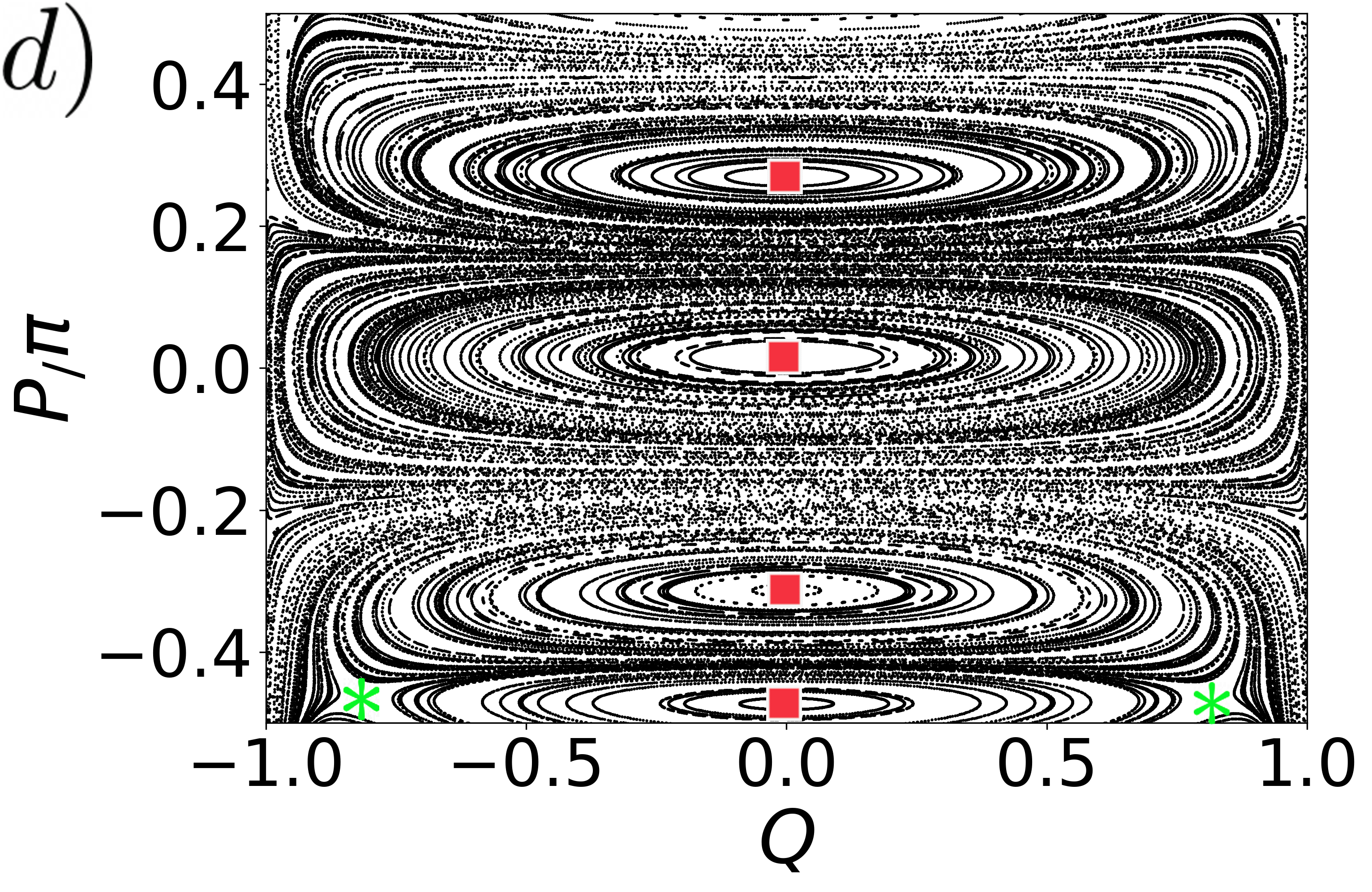} 
\includegraphics[scale=0.22]{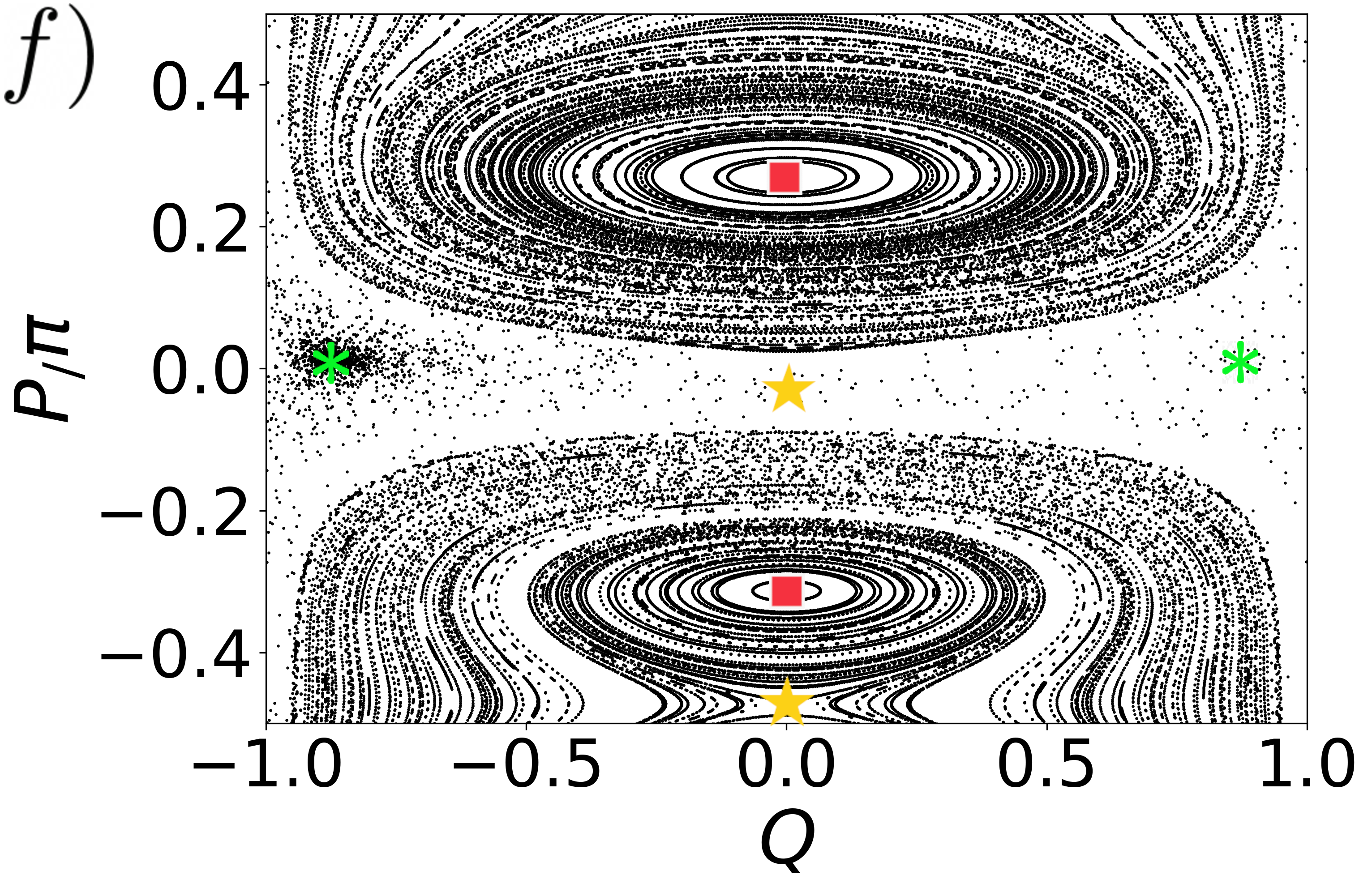}
 \caption{Phase space portraits $(Q,P)$ for the collective $2$-level system  with $\kappa = 0.5,\, \omega_0=1$ and different Hamiltonian couplings $\omega_x$ and  $\omega_z$.
  We show in (a-b) the results for a fixed 
  $\omega_x=0$ with 
 \textbf{(a)} $\omega_z=0$ and 
 \textbf{(b)} $\omega_z=2$. In
 (c-f) we show the phase portraits for the phase diagram line with fixed $\omega_x=2$, where 
 \textbf{(c)} $\omega_z=-1$,
 \textbf{(d)} $\omega_z=1$, 
 \textbf{(e)} $\omega_z=2$, 
 \textbf{(f)} $\omega_z=3$. 
 The red squares represent paramagnetic centers, the yellow stars are paramagnetic hyperbolic steady states and the green asterisks are pairs of ferromagnetic hyperbolic steady states.
 }
 \label{fig.phase.portrait}
\end{figure*}

We show in Fig.\eqref{fig.phase.diagram.fixed.points} the different phases of the model according to its steady state stability, for varying interactions $\omega_z$ and $\omega_x$ in the system Hamiltonian and considering both strong ($\kappa/\omega> 1$) and weak ($\kappa/\omega< 1$) dissipative cases.
We describe the phases by the $3$-vector $(n_{\rm pc}, \, n_{\rm ph},\, n_{\rm fh})$
where $n_{pc}$ denotes the number of paramagnetic steady states of type centers, thus associated to BTC periodic orbits, $n_{ph}$ the
number hyperbolic paramagnetic steady states and  $n_{\rm fh}$
the number of pairs of hyperbolic ferromagnetic steady states.
We see that the coherent interactions along the $x$-direction ($\omega_x$) tend to stabilize different center fixed points and consequently BTC's periodic orbits. On the other hand, interactions along the $z$-direction tend to generate pairs of ferromagnetic steady states or destroy the stability of BTC's paramagnetic steady states, turning their stability from centers to hyperbolic fixed points.  Interestingly, sufficiently high interactions $\omega_x$ can even lead to regions with the presence of fours centers in the model.
In Fig.\eqref{fig.phase.portrait} we show the phase portraits $(Q,P)$ defined by $m^z = Q$, $m^x = \sqrt{1 - Q^2}\cos(2P)$ and $m^y = \sqrt{1 - Q^2} \sin(2P)$ for a few points of the phase diagram, in order to make clearer our association of centers to BTC's in the model.

\section{Symmetry Breaking Perturbation}
\label{sec.symm.breaking.pert}

 In this section we study the effects of a specific perturbation in the system breaking the coherent  Hamiltonian $\mathbb{Z}_2$ symmetry. Motivated by the results of Sec.\eqref{sec.stability.analysis} showing that BTC's can only occur for paramagnetic steady states, we consider as perturbation a field along the $z$ - direction. In this way the model Hamiltonian is given by $\hat H_{\delta_z} = \hat H + \delta_z \hat S^z$ and the corresponding semiclassical dynamical equations are obtained:
\begin{eqnarray}
(\frac{d}{dt}m^x)_\delta &=& (\frac{d}{dt}m^x)_{\delta=0} -\delta m^y,\\
(\frac{d}{dt}m^y)_\delta &=& (\frac{d}{dt}m^y)_{\delta=0} + \delta m^x,\\
(\frac{d}{dt} m^z)_{\delta} &=& (\frac{d}{dt} m^z)_{\delta=0},
\end{eqnarray} 
where $(d m^\alpha /dt)_{\delta=0}$ denotes the unperturbed semiclassical equations (Eq.\eqref{eq:dynamical.equations.single}).
The Jacobian of the system can be written as 
\begin{equation}
\hat J_\delta = \hat J +\delta \hat V,
\label{eq:jacobian.plus.pert}
\end{equation}
 where $\hat J$ is the unperturbed Jacobian and 
\begin{equation}
 \hat V = \begin{pmatrix}
     0& -1 &0 \\
    1 &0  & 0\\
    0& 0 &0 
    \end{pmatrix},
\end{equation}
is the perturbation matrix. We can study the effects of such perturbation in the spectral properties of the Jacobian using Perturbation Theory for general matrices \cite{koch2014} (notice here that the Jacobian matrix is not an Hermitian matrix). The idea follows similarly to the simpler Hermitian case (see Appendix \eqref{sec.app.nonhermPT}).
In particular, the first order corrections to the eigenvalues ($\lambda_i^{(1)}$) are defined as 
\begin{equation}
 \lambda_i^{(1)} = 
\mathrm{Tr}( \vec{w}_i^{(0)^\dagger} \hat V \vec{u}_i^{(0)} )
\end{equation}
with $\vec{u}_i^{(0)}$ and $\vec{w}_i^{(0)}$ the left and right eigenvectors of the unperturbed Jacobian, respectively. We compute explictly these corrections for the paramagnetic steady states of our model (Eq.\eqref{eq:fixed.point.mz0.single}), and obtain
\begin{widetext}
\begin{align}
\lambda_{1}^{(1)} &= \frac{ −2(\kappa x_* − \omega_x y_*)(−\kappa y_* +\omega_0+2x_*(\omega_x−\omega_z))+(\kappa x_*−2\omega_z y_*)(−2\kappa y_* +\omega_0 + 2\omega_x x_*)}{2(\kappa x−\omega_x y_*)(\kappa x_*−\omega_z y_*)},
\nonumber \\
\lambda_{2}^{(1)} &=\frac{ 2(\kappa x_*− \omega_x y_*)(−\kappa y_* +\omega_0+2x_*(\omega_x−\omega_z))+(\kappa x_*−2\omega_z y_*)(−2\kappa y_* +\omega_0 + 2\omega_x x_*)}{2\kappa^2 x^2_* + 2\kappa^2 y^2_* - \kappa \omega_0 y_* - 8\kappa \omega_x x_*y_* +\omega_0^2 + 4\omega_0 \omega_x x_* -2\omega_0 \omega_z x_* + 4\omega_x^2 - 4\omega_x \omega_z x_*^2 + 4\omega_x \omega_z y_*^2}, \\
\lambda_{3}^{(1)} &=\frac{ 2(\kappa x_*− \omega_x y_*)(−\kappa y_* +\omega_0+2x_*(\omega_x−\omega_z))+(\kappa x_*−2\omega_z y_*)(−2\kappa y_* +\omega_0 + 2\omega_x x_*)}{2\kappa^2 x^2_* + 2\kappa^2 y^2_* - \kappa \omega_0 y_* - 8\kappa \omega_x x_*y_* +\omega_0^2 + 4\omega_0 \omega_x x_* -2\omega_0 \omega_z x_* + 4\omega_x^2 - 4\omega_x \omega_z x^2_* + 4\omega_x \omega_z y^2_*}. \nonumber 
\end{align}
\end{widetext}
Since $x^*,y^* \in \Re$ 
  the first order corrections are purely real terms, implying that the steady state centers become hyperbolic steady states. In this way, the closed orbits characteristics of BTC's are destroyed and we have  instead spirals towards or away from the fixed points of the model, with characteristic times captured by the real eigenvalues $\lambda_i^{(1)}$. The $Z_2$ Hamiltonian symmetry of the model is thus crucial for the stabilization of BTC's. 
  These results follow in accordance with closely related $p,q$- interacting model recently studied in Ref.\cite{giulia2021,PhysRevA.103.013306}, shown also to support BTC only in the absence of a $\mathbb{Z}_2$ Hamiltonian symmetry breaking perturbation.

\section{$d=4$: Dynamical Equations of Motion and Symmetries}
\label{sec.d4}

We move our studies now to the case of a pair of collective $2$-level ($1/2$-spin) systems. 
As in the previous case, the dynamical equations of motion can be obtained from a semiclassical approach.
Defining the operators $\hat m^\alpha_p = \hat S^\alpha_p/N$ and closing the expectations values in the second cumulant  $\langle \hat m_p^\alpha \hat m_p^\beta \rangle \cong \langle \hat m_p^\alpha\rangle \langle \hat m_p^\beta \rangle$ we obtain the semiclassical dynamical equations of motion: 
\begin{align}
\frac{d}{dt}m_1^x &= -\omega_{zz}m_1^ym_2^z - \omega_{z,1}m_1^y + \kappa_1m_1^xm_1^z, \nonumber \\
\frac{d}{dt}m_1^y &= \omega_{z,1}m_1^x - \omega_{x,1}m_1^z - \omega_{xx}m_1^zm_2^x +\omega_{zz}m_1^xm_2^z\nonumber \\
	&\quad+ \kappa_1m_1^ym_1^z, \nonumber \\
\frac{d}{dt}m_1^z &=\omega_{x,1}m_1^y +\omega_{xx}m_1^ym_2^x -\kappa_1\left((m_1^x)^2+(m_1^y)^2\right), \nonumber \\
\frac{d}{dt}m_2^x &= -\omega_{zz}m_1^zm_2^y - \omega_{z,2}m_2^y + \kappa_2m_2^xm_2^z,\\
\frac{d}{dt}m_2^y &= \omega_{z,2}m_2^x - \omega_{x,2}m_2^z - \omega_{xx}m_1^xm_2^z +\omega_{zz}m_1^zm_2^x\nonumber \\
	&\quad+   \kappa_2m_2^ym_2^z, \nonumber \\
\frac{d}{dt}m_2^z &= \omega_{x,2}m_2^y +\omega_{xx}m_1^xm_2^y -\kappa_2\left((m_2^x)^2+(m_2^y)^2\right). \nonumber
\label{eq:dynamical.equations.pair}
\end{align}

\textit{Symmetries and Conserved Quantities.- }
The dynamical equations conserve the total spin for each collective $1/2$-spin system $\mathcal{N}_p = (m^x)^2_p+(m^y)^2_p+(m^z)^2_p$  for $p=1,2$. 
We also notice that both couplings $\omega_{xx}, \omega_{zz}$ do not break the reversibility symmetry of the equations. 
In particular, in the case where $\omega_{z,1(2)} = \omega_{zz} = 0$ the equations still have the  factorization structure for the $m^z$ terms in the $m^x$ and $m^y$ dynamical equations. In this case one can proceed the analysis similarly to Ref.\cite{Iemini2018} and show that the system do have (quasi) conserved dynamical quantities.
In case  $\omega_{z,1(2)}$ or $\omega_{zz} \neq 0$, however, the coupling destroys this simpler factorization structure making inconclusive the existence of conserved quantities (one breaks also the $\mathbb{Z}_2$ Hamiltonian symmetry). 
 One could consider, however, the simpler case with only $\omega_{z,1(2)}=0$ and equal local couplings for both collective spins ($\omega_{x,1}=\omega_{x,2}$, $\kappa_{1}=\kappa_{2}$) and study the specific case where the collective spins are initially the same in the evolution. In this case they shall also remain the same  throughout all the dynamics, $\hat m^\alpha_1(t) = \hat m^\alpha_2(t)$ $\forall t$, and we recover the factorization structure in the dynamical equations, thus the existence of (quasi) conserved quantities.
 
\section{Route to Chaos}
\label{sec.route.to.chaos}

\begin{figure}
\includegraphics[scale=0.22]{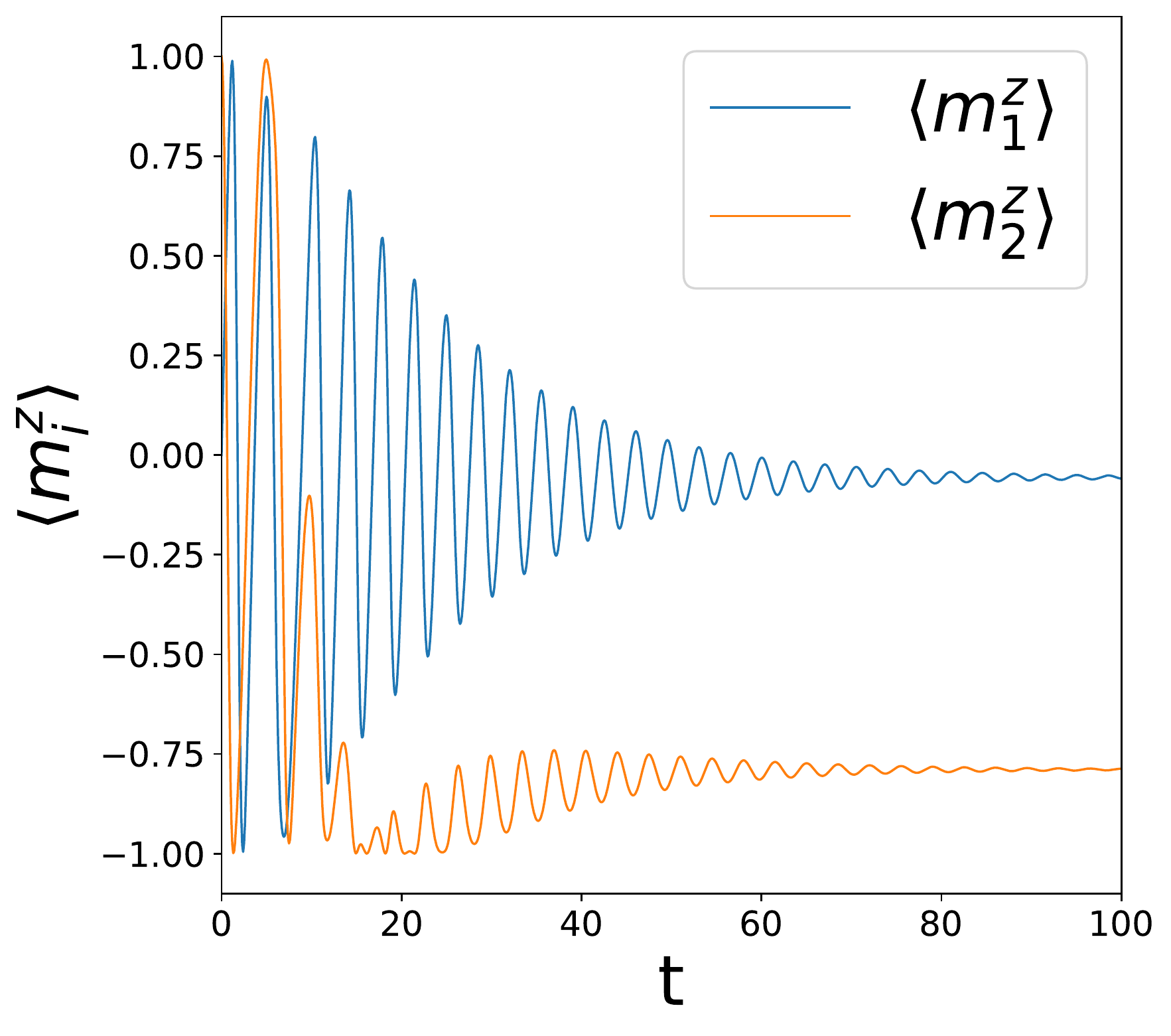}  \includegraphics[scale=0.22]{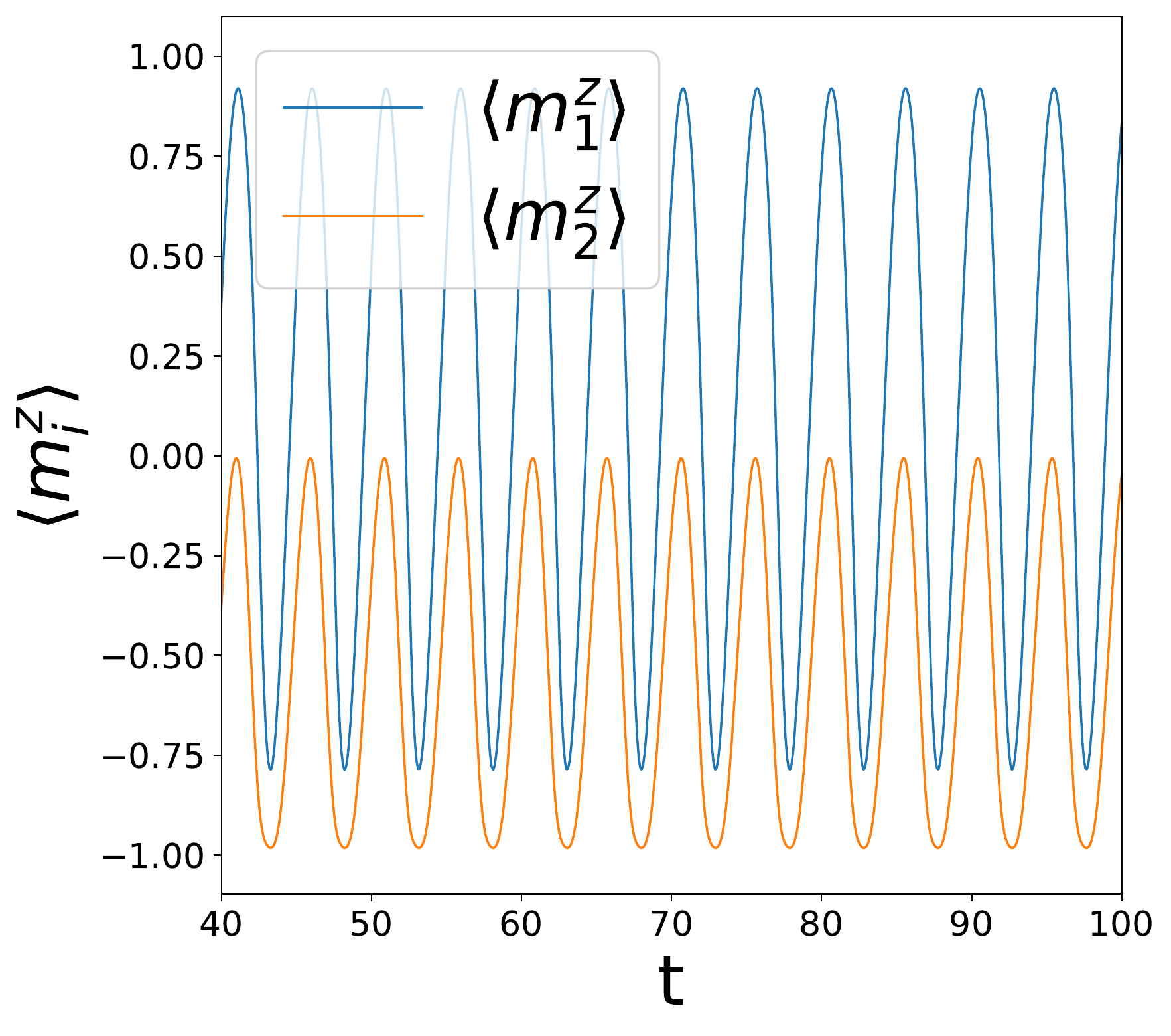}\\
\includegraphics[scale=0.22]{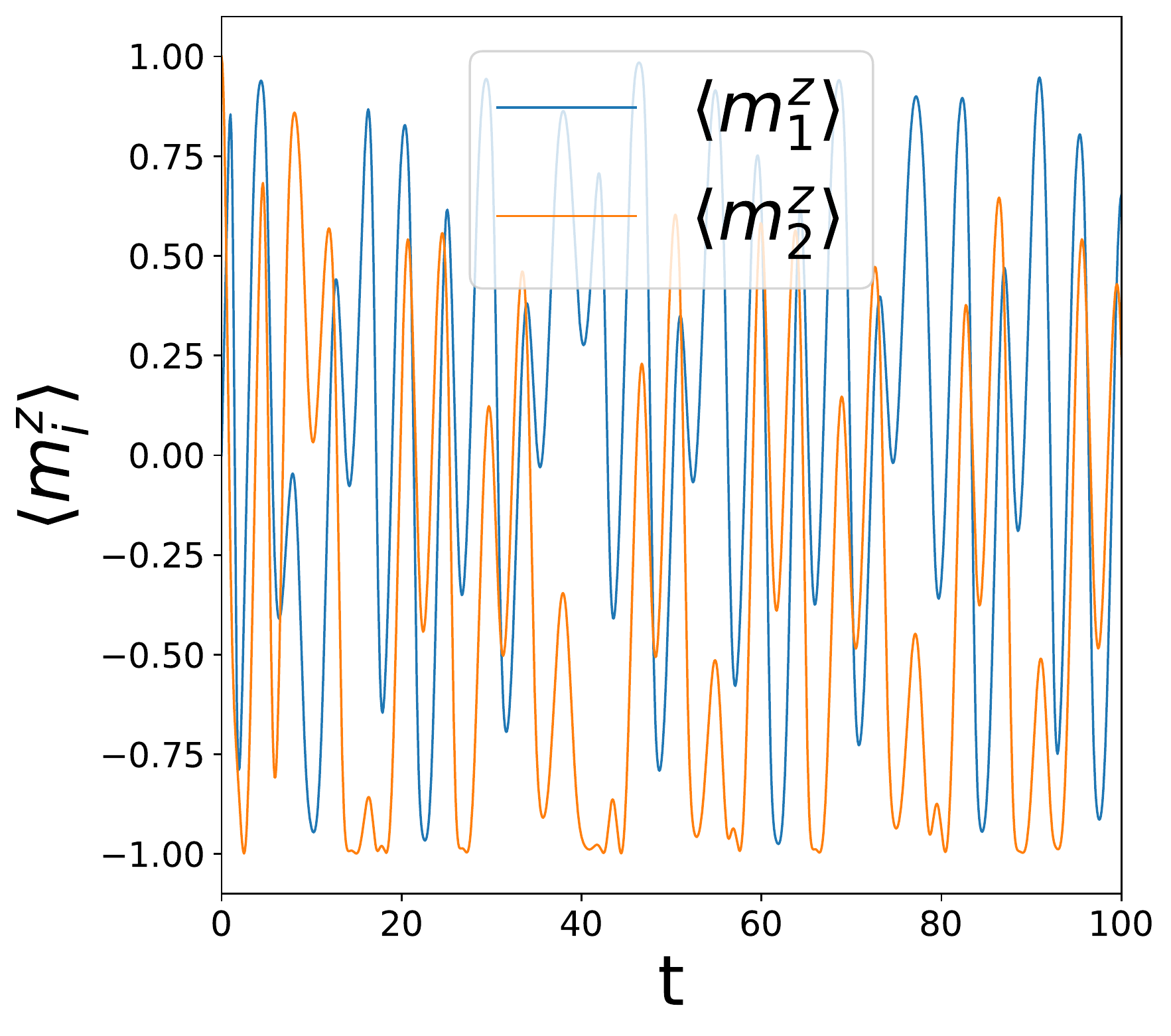} \includegraphics[scale=0.22]{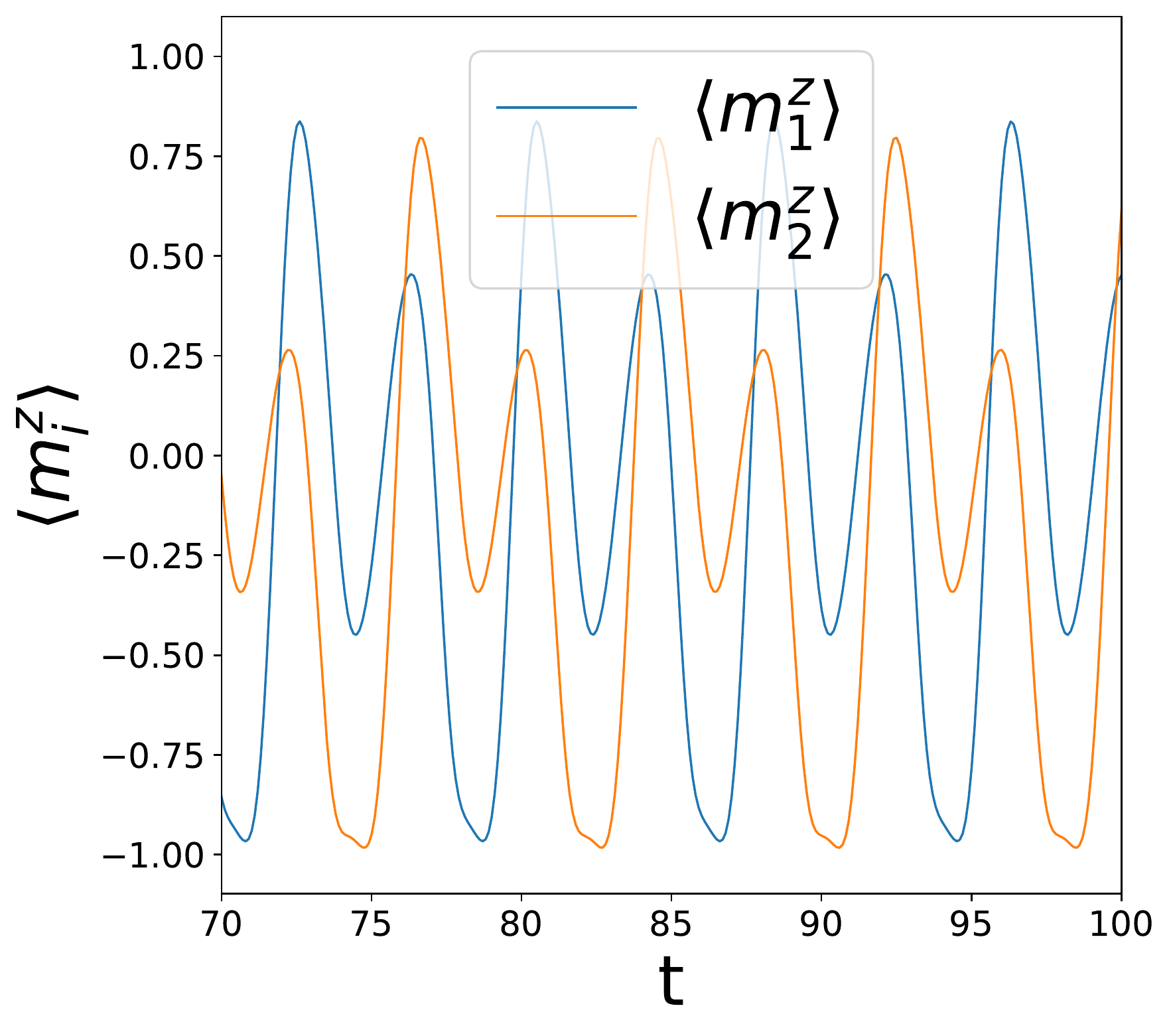}
 \caption{We show the dynamics of the spin magnetization along the z-axis for system parameters of Eq.\eqref{eq.system.param.pair} and different couplings $\omega_{zz}$ with $\vec{m}(0) = (0,1,0,0,0,1)$. The system supports different dynamical phases as one varies the coupling strength. We show in the \textbf{(upper-left panel)} a steady state with fixed magnetization for $\omega_{zz} = 0$, \textbf{(upper-right panel)} a BTC with limit-cycles for $\omega_{zz} = 0.35$, 
 \textbf{(bottom-left panel)} a chaotic dynamics for  $\omega_{zz} = 0.58$ and 
 \textbf{(bottom-right panel)} limit-cycles with period doubling oscillations for $\omega_{zz} = 1$.
 }
\label{fig.mag.dynamics.pair}
\end{figure}

\begin{figure}
\includegraphics[scale=0.21]{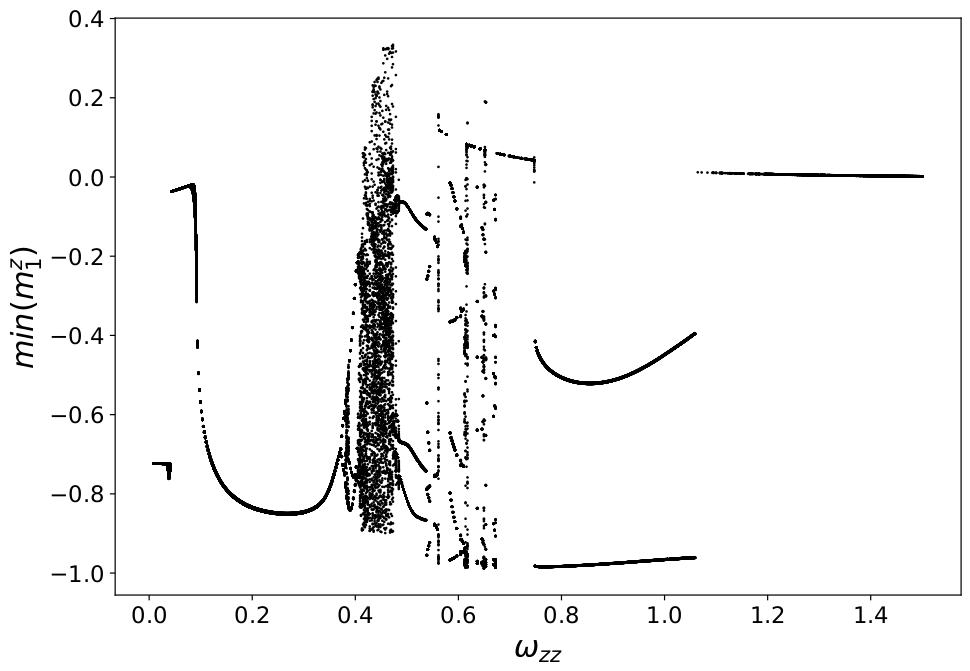}
\includegraphics[scale=0.21]{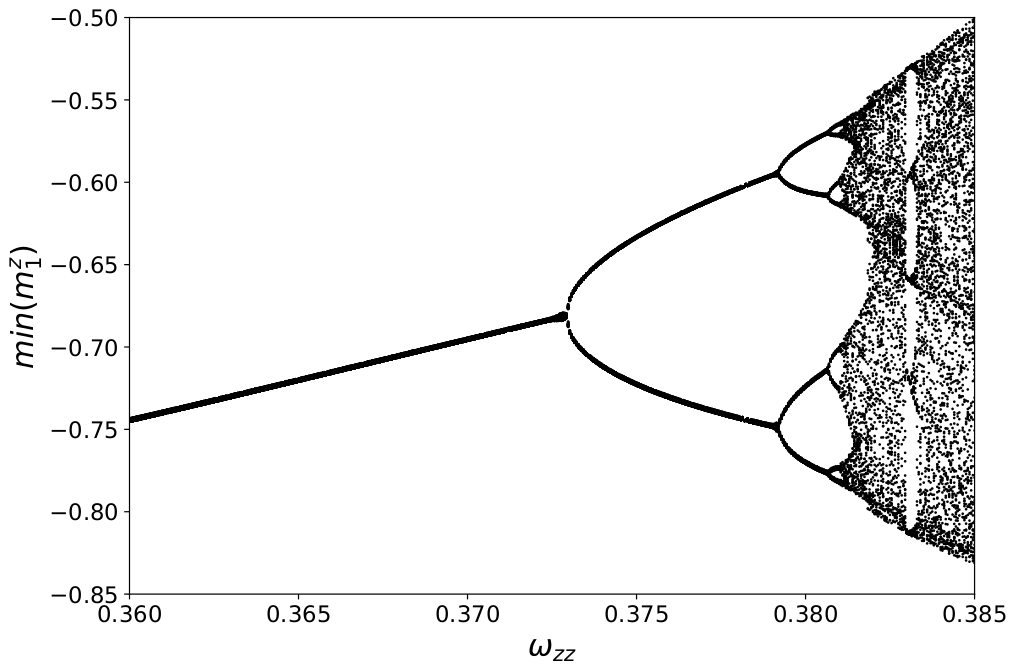}
 \caption{ \textbf{(top panel)} Orbit diagram for system parameters of Eq.\eqref{eq.system.param.pair} and varying $\omega_{zz}$. We show in the \textbf{bottom panel} a zoom for the region around $\omega_{zz} \approx 0.37$. The orbit diagram is obtained from the local minimums in the time
evolution of the system magnetization along the z-axis. We see a multitude of dynamical phases. For  
$\omega_{zz} \lesssim 0.1$ the system supports a ferromagnetic phase with nonzero steady state magnetization. As one increases $ 0.1 \lesssim \omega_{zz} \lesssim 0.4$ the system shows limit cycles and a period doubling cascade towards a chaotic dynamics. For $\omega_{zz} \sim 0.53$ we see the appearance of a $3$-cycle periodic window. For larger $\omega_{zz} \gtrsim 1$ the system stabilizes in a trivial (time independent) steady state with negligible magnetization. 
}
\label{fig.orbit.diagram}
\end{figure}

In the general case of nonzero couplings in the model the absence of (quasi) conserved quantities - beyond the norm $\mathcal{N}$ of the collective spins - and no $\mathbb{Z}_2$ Hamiltonian symmetry turns the analysis of the steady states and dynamics of the system more intricate. On the other hand, it allows the possibility of more complex dynamics with  richer dynamical phases. A particularly interesting case occurs when,
\begin{eqnarray}
&\mathrm{(i)} & w_{x,p}/\kappa_{p} > 1, \nonumber \\
&\mathrm{(ii)} & w_{z,p}/\kappa_{p} \ll 1,\nonumber  \\
&\mathrm{(iii)}& \omega_{xx}/\kappa_{p} \gg 1, \\
&\mathrm{(iv)} & \omega_{zz}/\kappa_{p} \neq 0,\nonumber  
\end{eqnarray}
for $p=1,2$. Conditions (i) and (ii) cannot alone support BTC's, as shown in the previous sections, rather they are characterized by hyperbolic steady states with $z_* \neq 0$. Condition (iv) can correlate these steady states with (iii) inducing coupled spin excitations on the system. The system in this case has no (quasi) conserved quantities, and the appearance of BTC's shall be due to the collective dynamics  of both spin systems (\textit{hybridization}).
 Specifically, we study the case with system parameters given by
\begin{eqnarray}
&\mathrm{(i)} & w_{x,p} = 2,\quad p=1,2,\nonumber \\
&\mathrm{(ii)} & w_{z,1} = 0.1, \quad  w_{z,2}  =  0.02, \nonumber  \\
&\mathrm{(iii)}& w_{xx} = 3,  
\label{eq.system.param.pair}
\end{eqnarray}
with $\kappa_{1}=\kappa_{2}=1$ and 
for varying coupling $\omega_{zz}$.
We show in Fig.\eqref{fig.mag.dynamics.pair} the dynamics for the magnetization along the $z$-axis for different cases of the coupling $\omega_{zz}$. We see that the system can support different phases ranging from (i) a ferromagnetic phase with nonzero magnetization for its steady states, $\langle \hat m^z_p (t \rightarrow \infty) \rangle  \neq 0$, (ii) BTC's characterized by limit-cycle oscillations, where after an initial transient time the magnetization oscillates in a given orbit indefinitely in time; moreover, further increasing the coupling the limit-cycles oscillations are followed by period doubling bifurcations till (iii) reaching a chaotic dynamics. Interestingly, we also see (iv) $3$-cycle periodic windows intercalated by the 
stable period doubling bifurcations and the chaotic regime. 

\begin{figure}
 \includegraphics[scale=0.27]{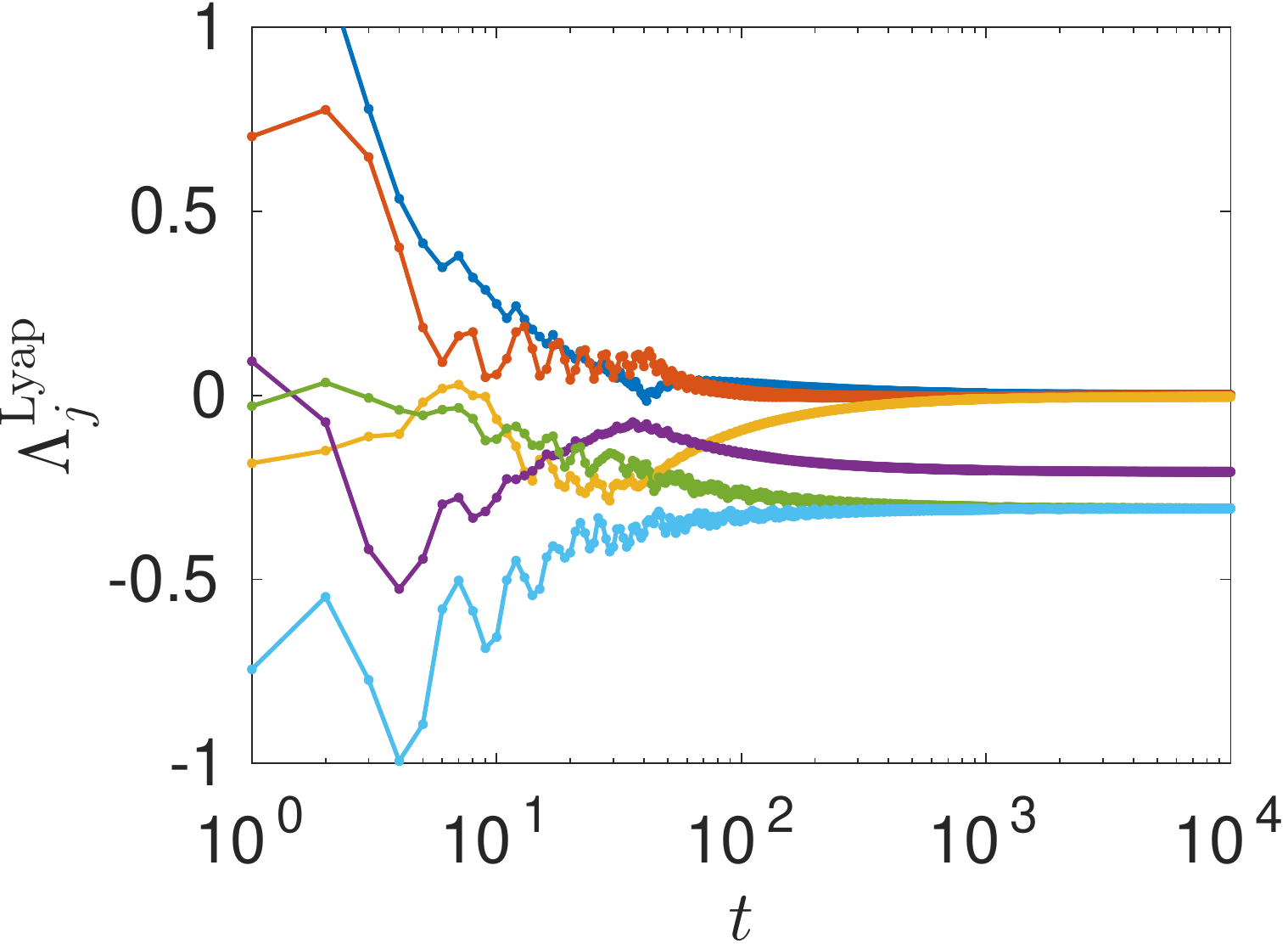}
 \includegraphics[scale=0.27]{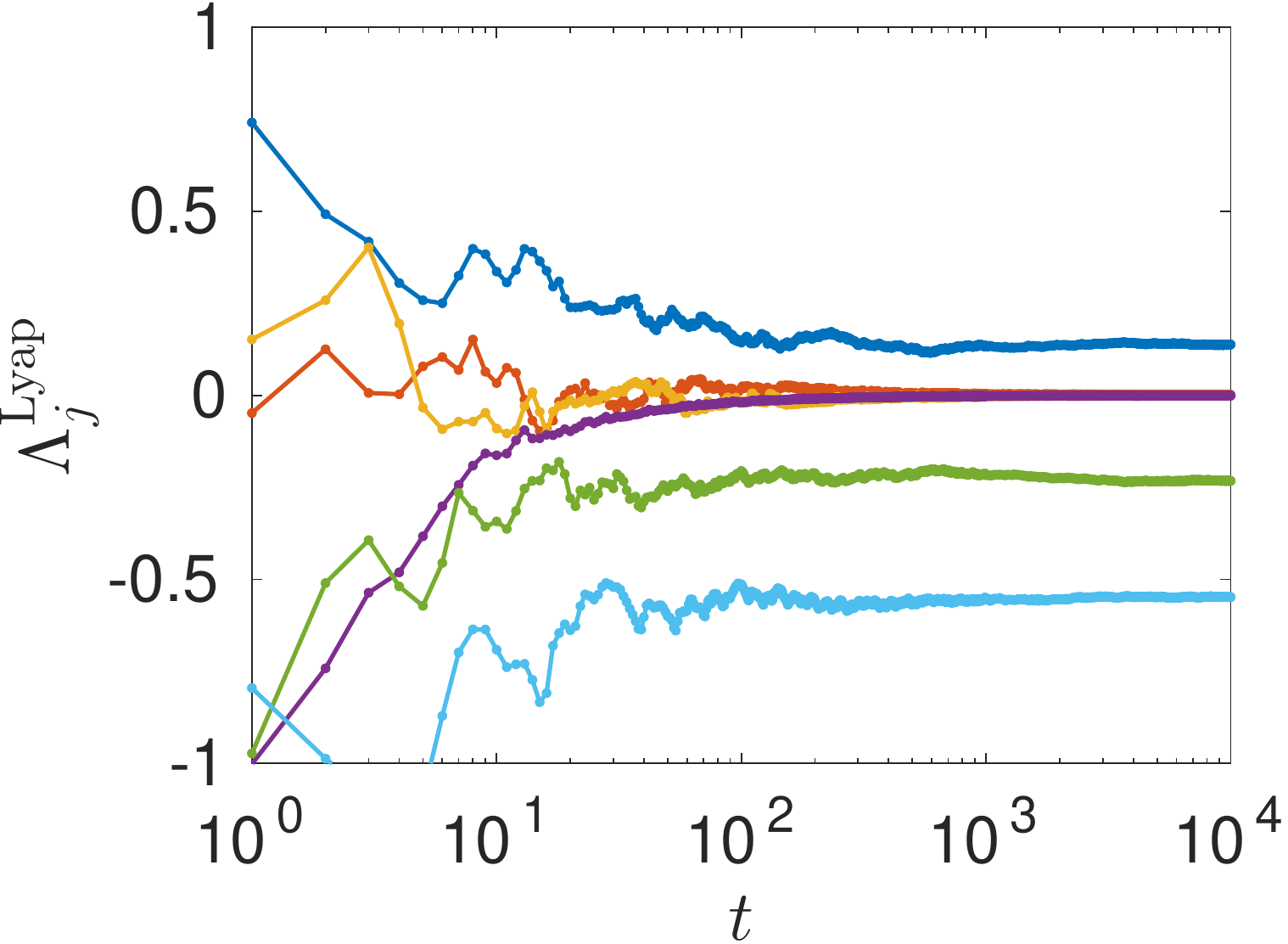}
 \caption{ We show the Lyapunov spectrum $\Lambda_j^{\rm lyap}$ obtained from Benettin's approach. We see the convergence of the $\Lambda_j^{\rm lyap}$ exponents towards its assymptotic value ($t \rightarrow \infty$). We also see the convergence of the assymptotic exponents as we decrease the time steps.  In particular we obtain that the largest Lyapunov exponent  for $\omega_{zz} = 0.35$ approaches zero in the long time limit as a power law with time, while for $\omega_{zz} = 0.45$ it approaches approximately $0.15$.  We used a time step $dt = 10^{-3}$ in our numerical simulations.
 }
\label{fig.liapunov}
\end{figure}

A global picture of the dynamical phases in the model are shown in the orbit diagram of Fig.\eqref{fig.orbit.diagram}. The orbit diagram corresponds to the local minimums in the time evolution of the magnetization along the z-axis, obtained after an  initial transient time. The transient time is related to the relaxation of the initial collective spin state towards the limit-cycle orbits, static steady states or chaotic regime.
 In our numerical simulations for the orbit diagram we used a fixed initial state for the dynamics, with $m_1^y(0) = m_2^z(0) = 1$ and zero otherwise. We observed however that the orbit diagram is qualitatively similar considering a few different initial conditions \cite{note1}.
 
We can accurately determine in the orbit diagram the first period doubling bifurcations as we increase the coupling $\omega_{zz}$. 
 We obtain a bifurcation ratio $b_3 \approx 4.2$, where 
 \begin{equation}
  b_n = \frac{ \omega_{zz}^{(n-1)} - \omega_{zz}^{(n-2)}}{ \omega_{zz}^{(n)} - \omega_{zz}^{(n-1)} },
 \end{equation}
with $\omega_{zz}^{(n)}$ the coupling corresponding to  the $n$'th period doubling bifurcation in the orbit diagram. Interesting to compare with Feigenbaum constant for the 
 seminal logistic map, in which one has $b_{n \rightarrow \infty} \approx 4.67$. In our model we obtained a slightly different value, which could indicate a different universality class for the period doubling cascades towards chaoticity. We remark, however, that we were able to obtain only the first bifurcation ratio $b_3$ (not precisely $n\rightarrow \infty$) so an irrefutable conclusion on the universality class cannot be drawn at this moment. It remains as a very interesting perspective for a future work.

We also study the Lyapunov exponents in the limit-cycle and chaotic regimes \cite{Benettin1976,benettin1980,Sandri1996,Pikovsky2015}. We obtain the full Lyapunov spectrum of the dynamics, describing the mean growth of an $n$-dimensional volume ($n=6$ in our case) in the tangent space. We use Benettin's approach in our analysis, i.e., employing recursively the Gram-Schmidt orthonormalization procedure during the stretching and folding of the $n$-dimensional volume. In our numerical simulations we set the same initial state as in the orbit diagram, and perform an initial transient evolution up to  $t=500$. After this initial evolution the collective spin is close to their corresponding dynamical phases, and we employ Bennetin's algorithm to obtain the Lyapunov spectrum. We show our results in Fig.\eqref{fig.liapunov}. We see that while for $\omega_{zz} = 0.35$ the spectrum is all non-positive, in the chaotic region with $\omega_{zz} = 0.45$ the largest Lyapunov exponent is positive, corroborating our orbit diagram expectations of a limit-cycle and chaotic regimes, respectively. We further notice that 
the sum of the Lyapunov spectrum is negative in both cases, typical of dissipative dynamical equations.

\section{$d=3$: Dynamical Equations of Motion and  Symmetries} 
\label{d3.dynamical.equations}

In this section we analyse the model with collective $d=3$-level system. In this case it is convenient to  work within the Gell-Man basis for the three-level subsystems,
\begin{eqnarray}\label{ Gell-Mann}
& \hat{g}^{1}_j = \left(\begin{matrix}
		0 & 1 & 0 \\\\
		1 & 0 & 0 \\\\
		0 & 0 & 0 \\\\
\end{matrix}\right),\, \hat{g}^{2}_j = \left(\begin{matrix}
		0 & -i & 0 \\\\
		i & 0 & 0 \\\\
		0 & 0 & 0 \\\\
\end{matrix}\right),\,\hat{g}^{3}_j = \left(\begin{matrix}
		1 & 0 & 0 \\\\
		0 & -1 & 0 \\\\
		0 & 0 & 0 \\\\
\end{matrix}\right), & \nonumber\\\nonumber\\
 & \hat{g}^{4}_j = \left(\begin{matrix}
		0 & 0 & 1 \\\\
		0 & 0 & 0 \\\\
		1 & 0 & 0 \\\\
\end{matrix}\right), \,
\hat{g}^{5} = \left(\begin{matrix}
		0 & 0 & -i \\\\
		0 & 0 & 0 \\\\
		i & 0 & 0 \\\\
\end{matrix}\right),\, \hat{g}^{6}_j = \left(\begin{matrix}
		0 & 0 & 0 \\\\
		0 & 0 & 1 \\\\
		0 & 1 & 0 \\\\
\end{matrix}\right), & \nonumber\\\nonumber\\   & \hat{g}^{7}_j = \left(\begin{matrix}
		0 & 0 & 0 \\\\
		0 & 0 & -i \\\\
		0 & i & 0 \\\\
\end{matrix}\right),\, \hat{g}^{8}_j = \frac{1}{\sqrt{3}}\left(\begin{matrix}
		1 & 0 & 0 \\\\
		0 & 1 & 0 \\\\
		0 & 0 & -2 \\\\
\end{matrix}\right) &
\end{eqnarray}
corresponding to an Hermitian basis for the $j$'th subsystem. The collective operators 
$\hat G^k =\frac{1}{2} \sum_{j=1}^N \hat g_j^k$ inherit directly the algebra of their microscopic constituents, \textit{i.e.} the SU($3$) algebra of the Gell-Man basis, given by 
$[\hat G^a,\hat G^b] = i\sum_{c=1}^8 f_{abc} \hat G^c,  $
with $a,b=1,...,8$ and $f_{abc}$ the structure constant totally
antisymmetric under the exchange of any pair of indices 
(see Appendix \eqref{app.gelmann}).
Any collective operator can be decomposed in this basis. Specifically, the coherent Hamiltonian terms of the model are decomposed as $\hat S_{12}^{x} = \hat G^1$,$\hat S_{23}^{x} = \hat G^6$, while the decay operators are given by 
   $ \hat S_{\pm,12} = \hat G^1 \pm i \hat G^2$ 
    and $\hat S_{\pm,23} = \hat G^6 \pm i \hat G^7$.

We can also define number operators for the three collective energy levels, corresponding to their collective occupation, as follows, 
\begin{eqnarray} \label{eq.number.op.su3}
 \hat N_1 &=& \frac{N}{3}\mathbb{I}+\hat G^3+\frac{1}{\sqrt{3}}\hat G^8, \nonumber \\
 \hat N_2 &=& \frac{N}{3} \mathbb{I} -\hat G^3+\frac{1}{\sqrt{3}}\hat G^8,  \\
 \hat N_3 &=&  \frac{N}{3} \mathbb{I} -\frac{2}{\sqrt{3}}\hat G^8,  \nonumber 
\end{eqnarray}
where $\mathbb{I}$ is the identity operator.
    
\textit{Dynamical equations of motion. }  Following  the same approach as in the previous sections, we define the operators $\hat m^k = \hat G^k/\left(N/2\right)$ and close the expectations values in the second cumulant  $\langle \hat m^k \hat m^\ell \rangle \cong \langle \hat m^k \rangle \langle \hat m^\ell \rangle$. We obtain the following  semiclassical dynamical equations of motion in the thermodynamic limit ($N \rightarrow \infty$):
\begin{widetext}
\begin{eqnarray}
\frac{d m^{1}}{dt}&=&k_{12}\left(1-\delta\right)m^{1}m^{3}+\frac{1}{2}\delta\left(\omega_{23}m^{5}-k_{23}\left(m^{4}m^{6}+m^{5}m^{7}\right)\right),\\
\frac{d m^{2}}{dt}&=&\left(1-\delta\right)\left(-\omega_{12}m^{3}+k_{12}m^{2}m^{3}\right)+\frac{1}{2}\delta\left(-\omega_{23}m^{4}+k_{23}\left(m^{4}m^{7}-m^{5}m^{6}\right)\right), \\
\frac{d m^{3}}{dt}&=& \left(1-\delta\right)\left(\omega_{12}m^{2}-k_{12}\left((m^{1})^{2}+(m^{2})^{2}\right)\right)+\frac{1}{2}\delta\left(-\omega_{23}m^{7}+k_{23}\left((m^{6})^{2}+(m^{7})^{2}\right)\right), \\
\frac{d m^{4}}{dt} &=&-\frac{1}{2}\left(1-\delta\right)\left(\omega_{12}m^{7}+k_{12}\left(m^{1}m^{6}-m^{2}m^{7}\right)\right)+\frac{1}{2}\delta\left(\omega_{23}m^{2}+k_{23}\left(m^{1}m^{6}-m^{2}m^{7}\right)\right), \\
\frac{d m^{5}}{dt}&=&\frac{1}{2}\left(1-\delta\right)\left(\omega_{12}m^{6}-k_{12}\left(m^{1}m^{7}+m^{2}m^{6}\right)\right)+\frac{1}{2}\delta\left(-\omega_{23}m^{1}+ k_{23}\left(m^{1}m^{7}+m^{2}m^{6}\right)\right), \\
\frac{d m^{6}}{dt}&=&\frac{1}{2}\left(1-\delta\right)\left(-\omega_{12}m^{5}+k_{12}\left(m^{1}m^{4}+m^{2}m^{5}\right)\right)+\frac{1}{2}\delta k_{23}\left(\sqrt{3}m^{6}m^{8}-m^{3}m^{6}\right), \\
\frac{d m^{7}}{dt}&=&\frac{1}{2}\left(1-\delta\right)\left(\omega_{12}m^{4}+k_{12}\left(m^{1}m^{5}-m^{2}m^{4}\right)\right)+\frac{1}{2}\delta\left(\omega_{23}\left(m^{3}-\sqrt{3}m^{8}\right)+ k_{23}\left(\sqrt{3}m^{7}m^{8}-m^{3}m^{7}\right)\right), \\
\frac{d m^{8}}{dt}&=&\frac{\sqrt{3}}{2}\delta\left(\omega_{23}m^{7}-k_{23}\left((m^{6})^{2}+(m^{7})^{2}\right)\right).
\end{eqnarray}
\end{widetext}

\begin{figure*}
\includegraphics[width = 0.85 \linewidth]{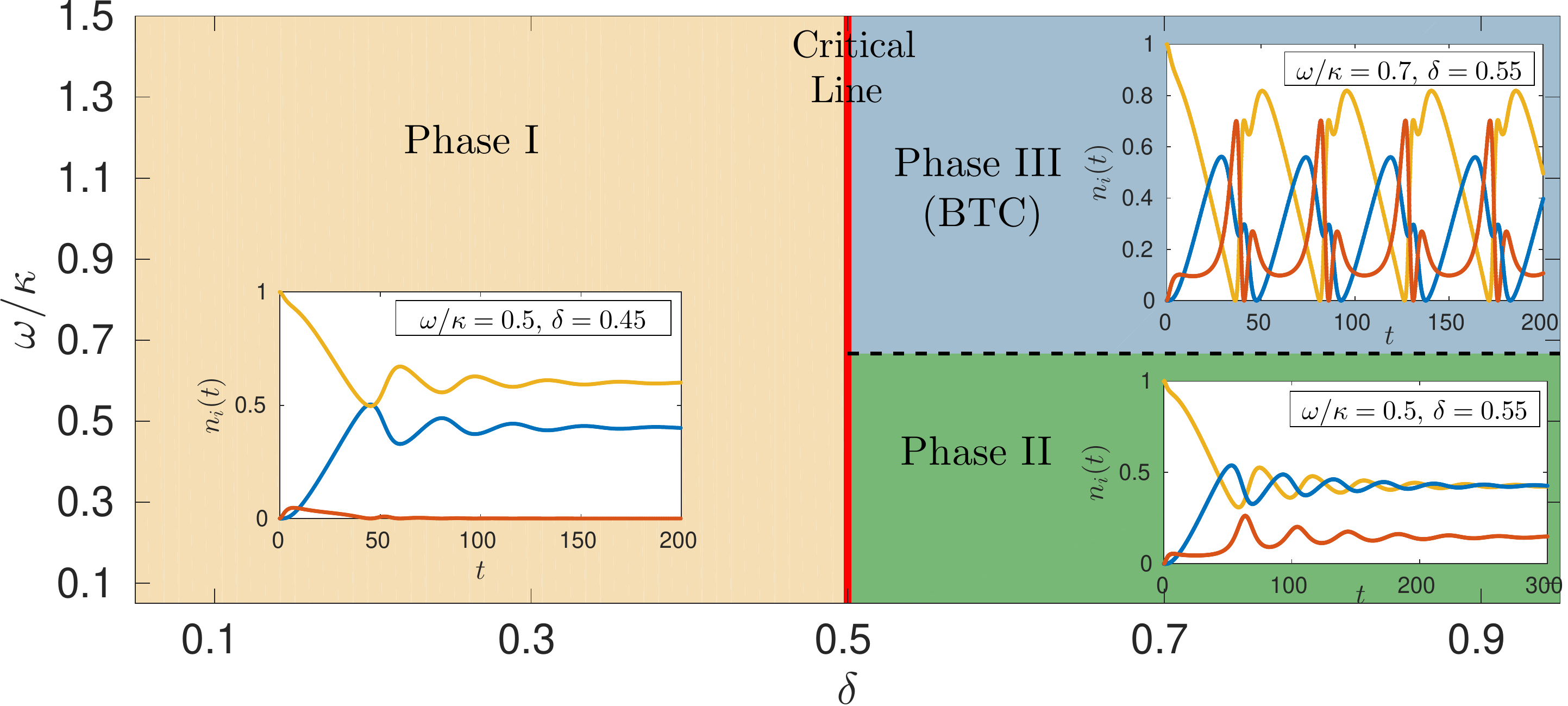}
\caption{ Phase diagram for the collective $3$-level system - Eqs.\eqref{eq.d3.Ham.Lind.definition.eq1},\eqref{eq.d3.Ham.Lind.definition.eq2} with couplings of Eq.\eqref{eq.couplings.d3}. We show three different phases supported by the model, phase I, II and III, with insets as illustrative of their dynamics (blue, red and yellow lines represent the dynamics of collective occupations $n_1(t)$, $n_2(t)$ and $n_3(t)$, respectively). The initial state for the dynamics is given by the collective occupation of a single level $n_3(t=0) = 1$. While phases I and II show static steady states with different characteristics, phase III features BTC's with limit-cycle dynamics. We highlight in red the critical line at $\delta =1/2$, supporting  multiple time crystal dynamical attractors, as discuss in the main text.
}.
\label{fig.phase.diag.su3}
\end{figure*}

\textit{Symmetries and conserved quantities. } 
 Since all collective operators can be decomposed in the $SU(3)$ basis, the model have conserved quantities given by the Casimir elements of the algebra (the element which commutes with all operators of the group). 
  The two independent Casimir elements in the  $SU(3)$ algebra correspond to a quadratic ($\hat C_2$) and a cubic operator ($\hat C_3$), defined as
\begin{eqnarray}\label{eq.casimir.su3}
\hat C_{2}&=&\sum_{j=1}^{8}\left(\hat{m}^{j}\right)^2,  \\
\hat C_{3}&=&\sum_{a,b,c}d_{abc}\hat{m}^{a}\hat{m}^{b}\hat{m}^{c},
\end{eqnarray}
respectively, where $d_{abc}$ is symmetric under the interchange of any pair of indices  (see Appendix \eqref{app.gelmann}).
We see that $C_2$ can be seen as a conservation of the norm in the basis of collective operators. This is similar to the $d=2$-dimensional case where the norm is associated to the total spin of the system, represented as  the surface of a $3$-dimensional sphere. In this case, however, the basis has a higher dimensionality, and the norm is related to the surface of a $8$-dimensional hypersphere. The interpretation of this surface with the the total spin of the system is not direct anymore.
This is also the case of the second Casimir element, $C_3$. It is cubic operator in the operator basis, further constraining the dynamics on the surface of the $8$-dimensional hypersphere. An exact interpretation of such constraint is, however, also not clear on physical grounds.

Moreover, even though each pair of energy levels in the subsystems are coupled similar to the $d=2$-dimensional case, the symmetries and conserved quantities present there are no longer present in this case, namely: reversibility (Eq.\eqref{eq.reversibility.sym.su2}), quasi-conserved quantities $\mathcal{R}$ (Eq.\eqref{eq.quasi.conserved.su2}) and total spin $\mathcal{N}$.

\section{$d=3$: Phase Diagram} 
\label{d3.dynamical.equations}

We study in this section the phase diagram for the $d=3$ model. We focus in the case of the Lindbladian with couplings 
\begin{equation}\label{eq.couplings.d3}
\omega_{12} = \alpha \omega_{23} \equiv \omega, \quad \kappa_{12} = \alpha \kappa_{23} \equiv \kappa,
\end{equation}
with $\alpha \in \Re$. We set $\alpha = 1$ in all of our analysis for simplicity, since different values corresponds simply to a renormalization of the Lindbladian and consequently its $\delta$ parameter. 
Specifically, for a different $\alpha$' coupling we see that $\mathcal{L}_{\alpha,\delta} = c \mathcal{L}_{\alpha',\delta'}$ where  $c  =\frac{\delta}{\delta '}$ and $\delta ' = \delta \alpha'/(\alpha (1-\delta) + \delta\alpha')$, thus both Lindbladians share the same steady states and dynamical attractors. 

The phase diagram is shown in Fig.\eqref{fig.phase.diag.su3}, characterized by different static as well as time crystal phases.
In all of our analysis we perform the dynamics starting from a few different initial state conditions and study its evolution towards their corresponding dynamical or static attractors. Precisely, we consider initial states with the collective system occupying the same level, i.e., with occupation number $\langle n_{i} (t=0) \rangle  = 1$ for $i=1,2$ or $3$, or initial states close to the steady state solutions of the model. We see no dependence of the attractors of the model on these considered initial conditions, except for the critical and extremal lines at $\delta =1/2$ and $\delta =0$ or $1$, respectively, as we will discuss in more detail. We remark however that since we deal with an $8$-dimensional space, we do not preclude the existence of different attractors in the model. A throughout analysis of the full set of steady states and their dependence on the initial conditions remains as an interesting perspective.
We discuss in detail the different phases of model below. 

$\bullet$ Extremal lines: for couplings $\delta =0$ or $1$ one recovers the $SU(2)$ model of Sec.\eqref{d2.dynamical.equations} for the pair of energy levels $(1,2)$ or $(2,3)$, respectively. In this case only a pair of levels has nontrivial dynamics, supporting a ferromagnetic or BTC phase, depending on the strength $\omega/\kappa$.

\begin{figure}
 \includegraphics[scale=0.31]{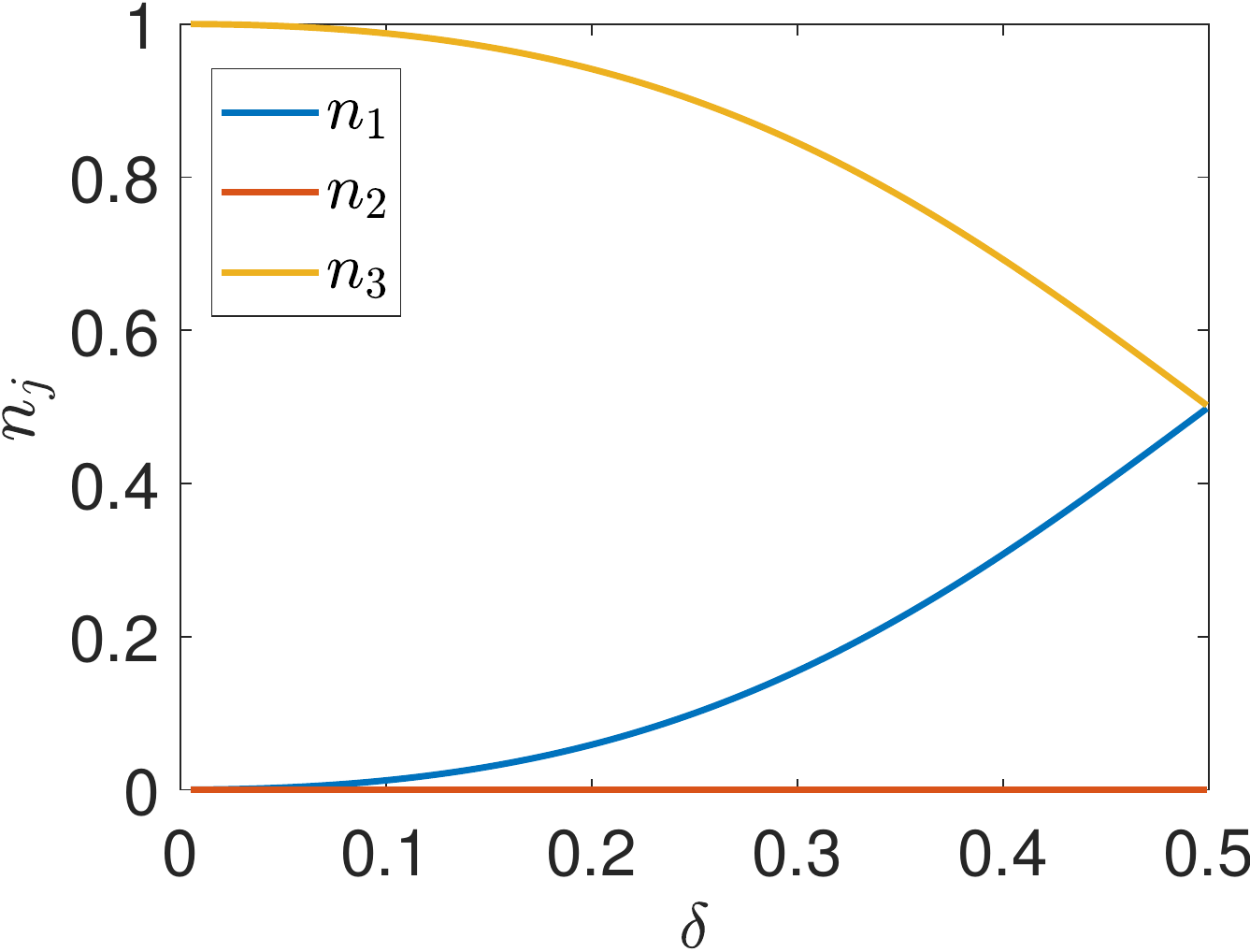}
 \includegraphics[scale=0.31]{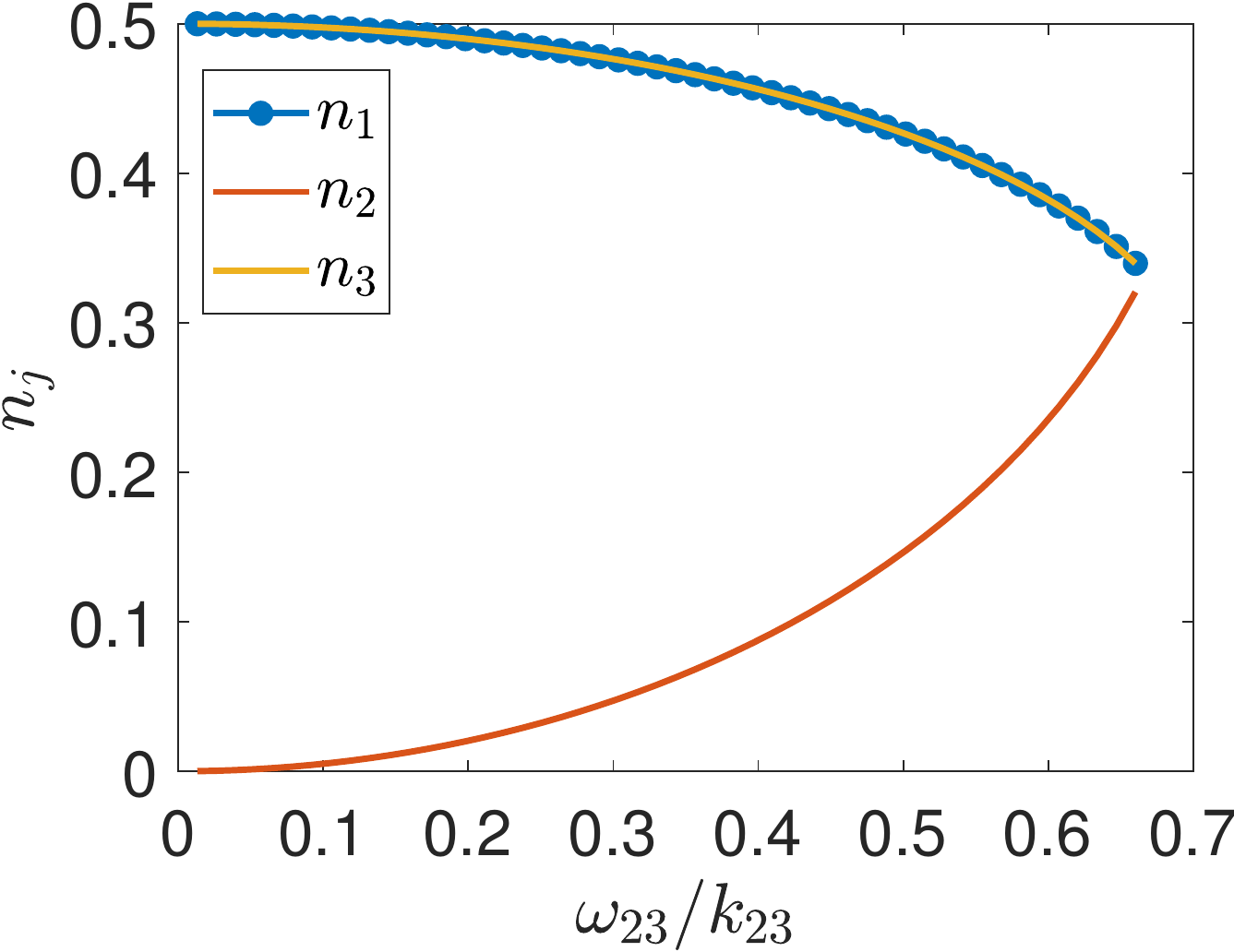}
\caption{We show the occupation numbers for the three collective energy levels of the system: (left panel) steady states of Phase I - Eq.\eqref{ss.su3.delta.less.0.5} - for $0<\delta <1/2,$ $\forall \omega/\kappa$; (right panel) steady states of Phase II - Eq.\eqref{ss.su3.delta.more.0.5} - for $ 1/2 < \delta < 1$ and $\omega/\kappa <2/3$.}
\label{fig.trivial.phases}
\end{figure}

$\bullet$ Phase I: for couplings $0<\delta < 1/2$, the Lindbaldian acts stronger on levels $(1,2)$.  The presence of the competing Lindbladian $\mathcal{L}_{2,3}$, even if small, tends to destroy the $SU(2)$ organization on these levels. We obtain the following steady state attractor for the model,
\begin{eqnarray}\label{ss.su3.delta.less.0.5}
 m^{3} &=& \frac{\sqrt{3C_2}}{2}\frac{\delta^2}{2\delta^2-2\delta+1} \nonumber \\
 m^4 &=& -2 \frac{ (1-\delta) }{\delta} m^3 \nonumber \\
 m^8 &=& \frac{1}{\sqrt{3}}(1 - 2 \frac{(1-\delta)^2 }{\delta^2}  ) m^3 \nonumber \\
 m^{i} &=& 0,\qquad i=1,2,5,6,7.
\label{eq.fixed.su3.a}\end{eqnarray}
where $C_2 = 4/3$ is the conserved quantity (quadratic Casimir element). We show in Fig.\eqref{fig.trivial.phases} the occupation numbers on such a phase. The energy level $2$ tends to be suppressed in the dynamics and becomes a ``dark mode'' in the steady state. Not only its number occupation is null, as well as there are no coherence between it and other energy levels ($m^{i} = 0,\, i=1,2,6,7$). 
The steady states do not depend on the ratio $\omega/\kappa$ and the phase transition at $\delta=1/2$ occurs when occupation numbers $n_1$ and $n_3$ equilibrate.

\begin{figure}
 \includegraphics[scale=0.25]{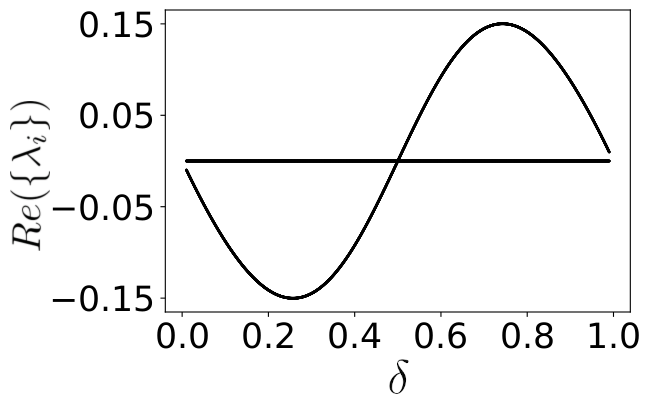}
 \includegraphics[scale=0.25]{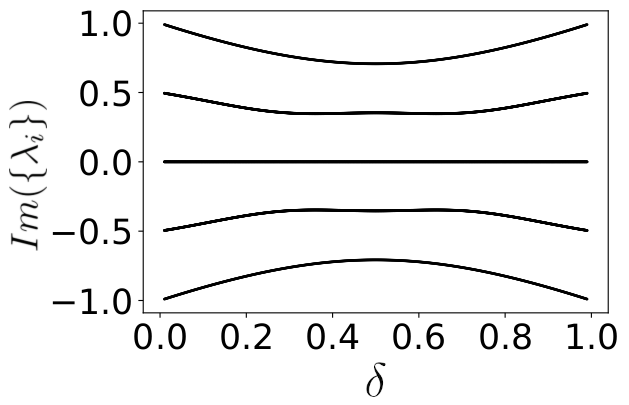}
 \includegraphics[scale=0.25]{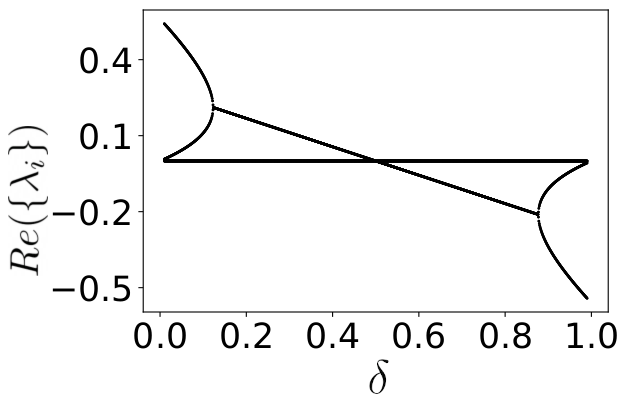}
 \includegraphics[scale=0.25]{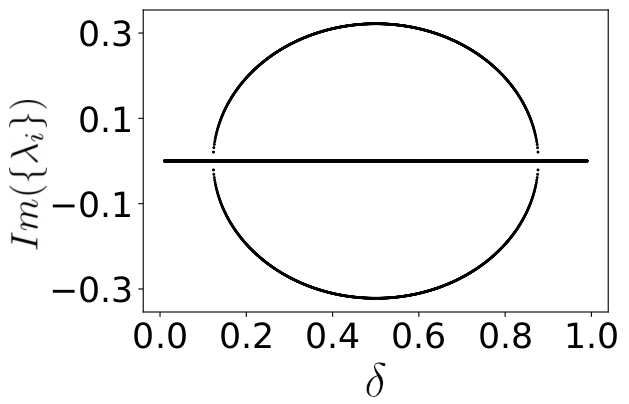}
\caption{Jacobian spectrum for the steady state of phases I and II along the line with $\omega/\kappa=1/2$ for varying $\delta$. In the upper panels we show the real and imaginary terms of the Jacobian eigenvalues for the steady states of phase I - Eq.\eqref{ss.su3.delta.less.0.5}
In the bottom panel we show the spectrum for phase II - Eq.\eqref{ss.su3.delta.more.0.5}.
Except for $\delta =1/2$, in both cases the spectrum has a four-fold degeneracy of eigenvalues with zero real part.}
\label{fig.jacobian.trivial.phases}
\end{figure}

The conditions of Eq.\eqref{ss.su3.delta.less.0.5} is actually valid as a steady state  $\forall \delta$, i.e., it is a solution of the dynamical equations. However, depending on $\delta$ it does not characterizes the dynamics of the system, since it becomes an unstable fixed point. In order to highligh it we show in Fig.\eqref{fig.jacobian.trivial.phases}-(upper panels) the Jacobian spectrum for such steady states. While it is an attractor for $\delta <1/2$, it becomes a repulsor for $\delta > 1/2$ (or an unstable steady state). At the transition point $\delta=1/2$ the full spectrum has only imaginary terms.

$\bullet$ Phase II: for couplings $1/2<\delta <1$ and  $\omega/\kappa < 2/3$, we obtain the
following steady state attractor,
\begin{eqnarray}\label{ss.su3.delta.more.0.5}
m^2 &=& \omega/\kappa \nonumber \\
 m^{3} &=& (-1 + 3 \sqrt{ 1-2 (\omega/\kappa)^2 })/4
  \nonumber \\
 m^{4} &= & - \sqrt{C_2 -2 (\omega/\kappa)^2 - \frac{4}{3} (m^3)^2 }
  \nonumber \\
 m^7 &=& m^2 \nonumber \\
 m^8 &=& -m^3/\sqrt{3} \nonumber \\
 m^i &=& 0, \qquad i=1,5,6.
\label{eq.fixed.su3.b}\end{eqnarray}
We show in Fig.\eqref{fig.trivial.phases} the occupation numbers along this phase. The steady states do not depend on the coupling $\delta$ and the occupation numbers for energy levels $1$ and $3$ are now equal,  with a nonzero occupation for energy level $2$. 

Similar to phase I, the conditions of Eq.\eqref{ss.su3.delta.more.0.5} is valid as a steady state for a larger range in the phase diagram, specifically, it is a solution of the dynamical equations $\forall \delta $ with $\omega/\kappa < 2/3$, though not always stable.
In Fig.\eqref{fig.jacobian.trivial.phases}-(bottom  panels) we show the Jacobian spectrum for such steady states. We see that it corresponds to a repulsor (unstable steady state) for  $\delta <1/2$, while becoming an attractor for $\delta > 1/2$. At the transition point $\delta=1/2$ the full spectrum also has only imaginary terms.

$\bullet$ Phase III: for couplings $1/2<\delta < 1$ and $\omega/\kappa \gtrapprox 2/3$ we observe a boundary time crystal with the appearance of limit-cycles in the dynamics. We show in Fig.\eqref{fig.phase.diag.su3}-(inset) the dynamics of occupation numbers for an  illustrative case. We remark that for different initial states, we observed the same dynamical limit-cycle attractor. Interesting to notice that the time crystal occurs not only in the weak dissipative regime ($\omega/\kappa > 1$), as in the $SU(2)$ case, but also in a region of the strong dissipative regime ($2/3 \lessapprox \omega/\kappa < 1$), indicating greater robustness of the phase.

\begin{figure}
\includegraphics[scale=0.27]{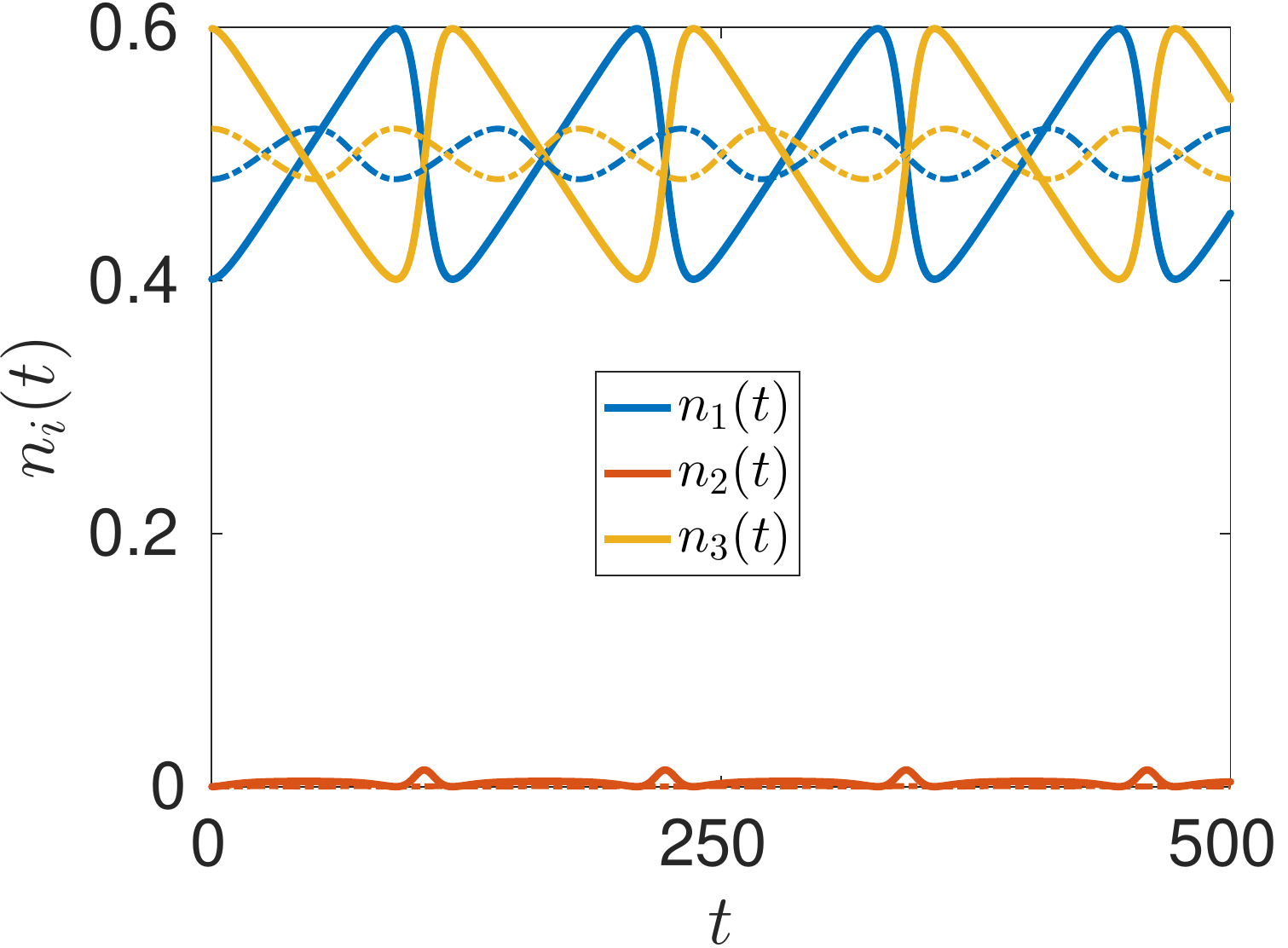} 
\includegraphics[scale=0.27]{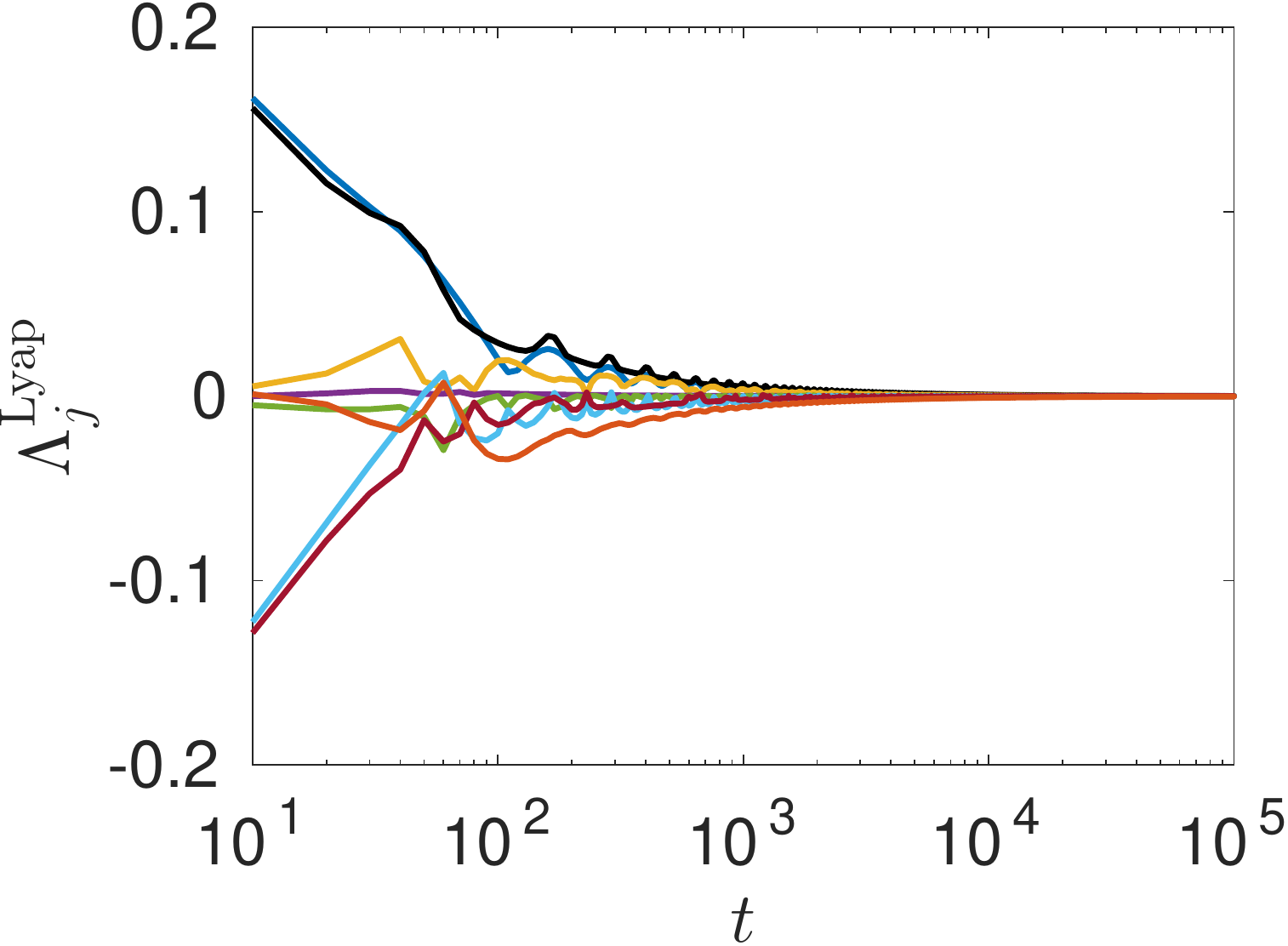}
\includegraphics[scale=0.27]{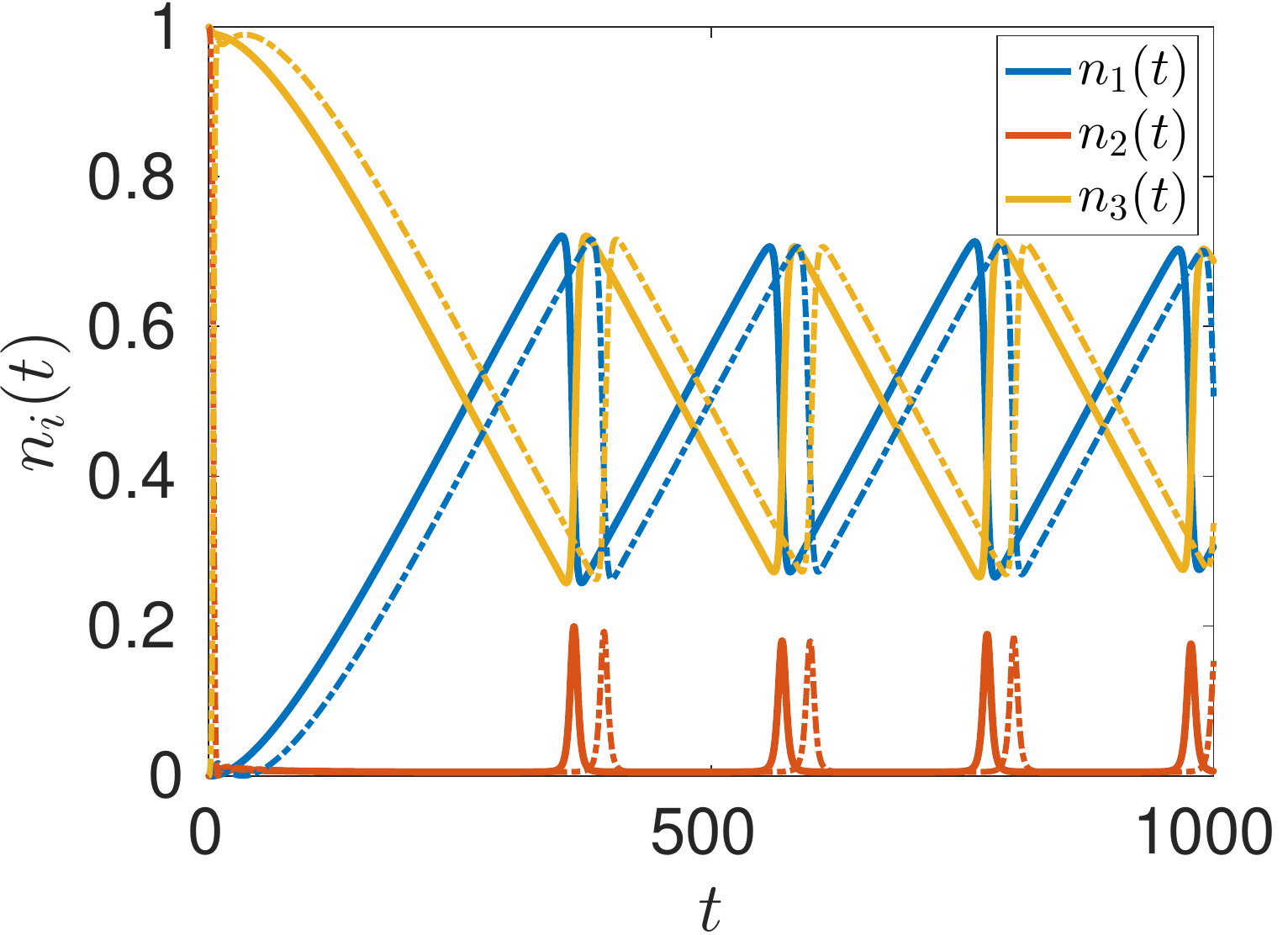} 
\includegraphics[scale=0.27]{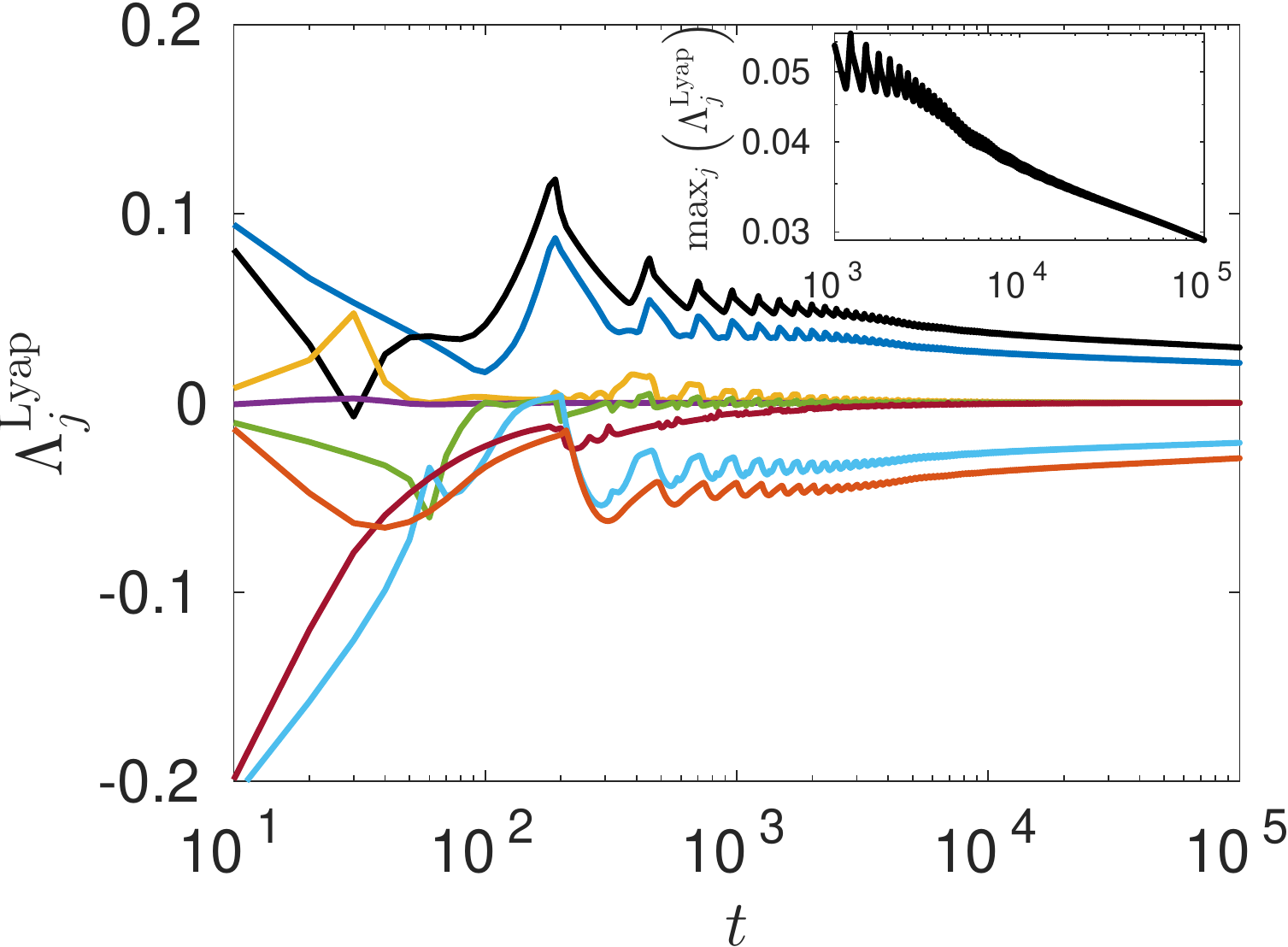}
\caption{ \textbf{(Left panels)} We show the dynamics of the occuppation number for different initial conditions, with system parameters $\delta = 1/2$ and $\omega/\kappa = 0.2$: (upper left) for initial states with $n_3(0) = 1$ (continuous curves) and $n_{2}(0) = 1$ (dotted curves); (bottom left) for initial states given by Eq.\eqref{ss.su3.delta.less.0.5} with $\delta = 0.45$ (continuous curves) and $\delta = 0.49$ (dotted curves).
\textbf{(Right panels)} Lyapunov spectrum obtained from Benettin's approach for the same system parameters: (upper right) using initial state $n_3(0) = 1$; (bottom right) for initial state given by Eq.\eqref{ss.su3.delta.less.0.5} with $\delta = 0.45$. We show in the inset the largest Lyapunov exponent decaying for long times, in a log-log scale.
In both cases we used a
time step $dt = 10^{-3}$ for our numerical simulation.
}.
\label{fig.dyn.lyap.critic.delta}
\end{figure}

$\bullet$ Critical line: for coupling $\delta=1/2$ and $\forall \omega/\kappa$ we observe a very peculiar behavior. As discussed, the steady states of Eqs.\eqref{ss.su3.delta.less.0.5} are solutions of the dynamical equations also for $\delta=1/2$. In this case, however, its Jacobian spectrum has only imaginary terms, corresponding to center steady states. It indicates that for a given initial state close to these steady states, the dynamics should correspond to closed orbits, typical of the BTC's found in the collective $d=2$ system. We find that this is indeed the case. We show in Fig.\eqref{fig.dyn.lyap.critic.delta}-(upper left panel) the dynamics for two initial states close to the steady states of Eq.\eqref{ss.su3.delta.less.0.5} for $\delta=1/2$. Both show closed, but different, orbits.

Considering the case of an initial condition far from these steady states, thus out of a Jacobian linear stability approach, we see now the existence of limit-cycles. We show in Fig.\eqref{fig.dyn.lyap.critic.delta}-(bottom left panel) the dynamics for initial states with $n_i(0) = 1$, for $i=3$ or $2$, both featuring the same limit-cycle dynamics. We also observed that other different initial states may lead to even further different limit-cycles.
We remark that for larger coherent coupling $\omega/\kappa \gtrapprox 1/2$ the limit-cycles are not so apparent anymore, mostly resembling as closed period orbits - there is no clear transient time in the dynamics towards the dynamical attractor. In this case it is not conclusive the existence of limit cycles, however it is still clear the presence of multiple time crystal dynamical attractors.

We study also the Lyapunov spectrum for this critical line, see Fig.\eqref{fig.dyn.lyap.critic.delta}-(right panels). We obtain that the full Lyapunov spectrum for all these different initial conditions is zero, corroborating the expectation of multiple BTC's.

\section{Conclusion}
\label{sec.conclusion}

In this work we studied boundary time crystals in collective $d=2,$ $3$ and $4$ level systems. We obtained that the BTC phase can appear in different forms for these different cases, highlighting a richer phenomenology for  such dynamical phases.
We first considered the model with $d=2$ presented in \cite{Iemini2018} and extended the analysis of its phase diagram, obtaining the full set of steady states combining  analytical and algebraically (quasi-analytically) approaches, and further studying its Jacobian stability. The existence of BTC in the model is seen to be directly related to the presence of \textit{center} fixed points. Moreover, we obtained analytically the effects of a specific $\mathbb{Z}_2$ symmetry breaking Hamiltonian term to the dynamics, showing that BTC's are destroyed by such a perturbation in the model. 

The analysis of the collective $d=4$-level system, composed of a pair of collective interacting $2$-level systems, showed even more fruitful. The model supports more robust forms of BTC's, from limit-cycles to period doubling bifurcations leading to chaos. The BTC is robust to $Z_2$ symmetry breaking Hamiltonian term in this case. We obtained the orbit diagram of the model from its collective magnetization and extracted its bifurcation ratio $b_n$ for a finite $n$. The bifurcation ratio for finite $n$ was found different from the Feigenbaum constant of the seminal logistic map. A careful analysis for the ratio in the limit $n \rightarrow \infty$ remains as an interesting perspective for a future work.

In the collective $d=3$-level system we observed that depending on the competition between the two Lindbaldians $\mathcal{L}_{12}$ and $\mathcal{L}_{23}$ the model supports static steady states characterized by a “dark level”,
limit-cycle dynamics or a peculiar dynamical phase at
the critical line $\delta = 1/2$ where the strength of both channels are equal. At this critical line the systems shows multiple limit-cycle attractors, depending on the initial condition, as well as different closed period orbits. The Jacobian spectrum has no real terms (only nontrivial imaginary part) in this phase, as well as its  Lyapunov exponents are all zero.

This work opens different interesting perspectives, showing how collective models with high $d$-level systems can support different forms of BTC's with richer properties. It would be interesting \textit{e.g.} to explore larger $d$'s and its implications to these phases, possible applications as well as a throughout analysis on the role of the global and dynamical symmetries for such models. 

\section*{Acknowledgements}
We acknowledge enlightening discussions with R. Fazio. F.I. acknowledges the financial support of the Brazilian funding agencies National Council for Scientific and Technological Development—CNPq (Grant No. $308205/2019$-$7$) and
FAPERJ (Grant No. E-$26/211.318/2019$).

\appendix

\section{Paramagnetic steady states ($d=2$)}
\label{sec.append.d2Paramagnetic}

\begin{figure}
 \includegraphics[scale=0.4]{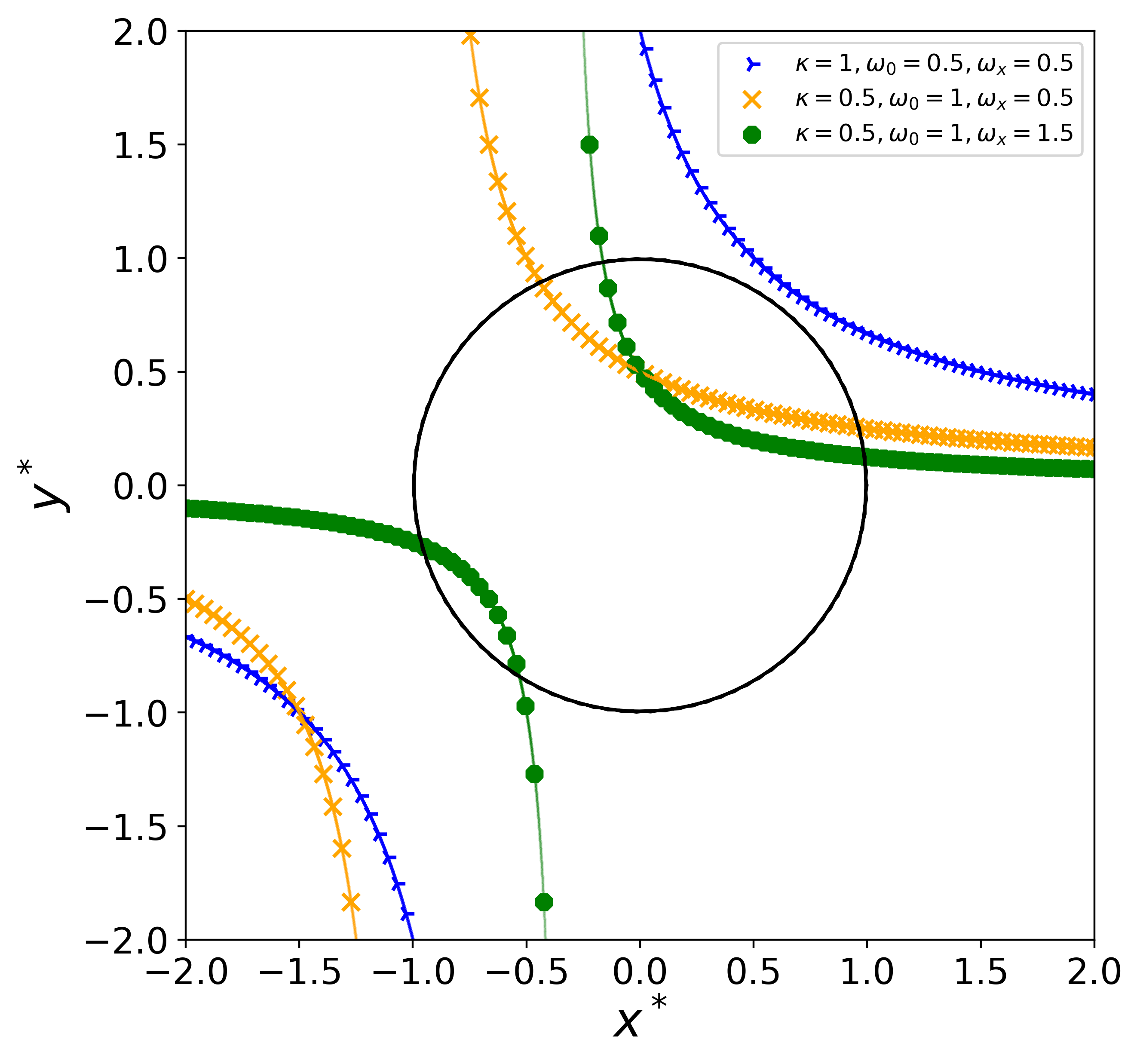} 
 \caption{Steady states of the model with $m^z=0$:   The colored curves are the algebraic condition of Eq.\eqref{eq:fixed.point.mz0.single} while the black circle the normalization condition $(m^x)^2 +  (m^y)^2+ (m^z)^2=1$. The physical steady states correspond to the intersection of the two curves. We show different cases of system parameters, highlighting the cases without such steady states, a pair and two pairs.
}
 \label{fig.fixed.points.mz0.intersections}
\end{figure}

\begin{figure}
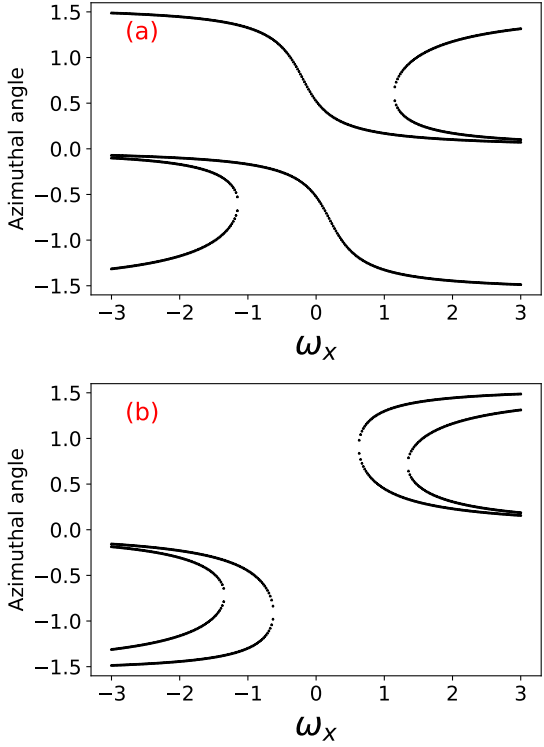

\includegraphics[width = 0.85 \linewidth]{azimuth_weak.png}
\includegraphics[width = 0.85 \linewidth]{azimuth_strong.png}
\caption{Azimuthal angle for the steady states with $m^z=0$ (Eq.\eqref{eq:fixed.point.mz0.single}), for varying $\omega_x$ field and system parameters \textbf{(a)} $\kappa=0.5, \omega_0=1$, and 
\textbf{(b)} $\kappa=1, \omega_0=0.5$. The steady states are independent of  the $\omega_z$ field.
}
 \label{fig.azimuth}
\end{figure}

The paramagnetic fixed points of the $2$-level collective model in Eqs.\eqref{eq.d2.Ham.Lind.definition.eq1}-\eqref{eq.d2.Ham.Lind.definition.eq2} are given by the algebraic condition,
\begin{equation}
y^* = \frac{\kappa}{\omega_0 + 2\omega_x x^*},
\label{eq:fixed.point.mz0.single}
\end{equation}
and we recall that this fixed point solutions are physical steady state only if this geometrical place intersects the fixed norm circle $(m^x)^2 + (m^y)^2 = \mathcal{N}$, thus satisfying the conservation of the total spin. 
In this case the steady states come in pairs, as in the previous case, however we can have $0,1$ or $2$ pairs depending on the system parameters $\kappa,\omega_0$ and $\omega_x$ - see Fig.\eqref{fig.fixed.points.mz0.intersections} for illustrative  cases.
We also show in Fig.\eqref{fig.azimuth} the steady states magnetization for varying Hamiltonian parameters. Since $m^z=0$ we use the azimutal angle in the $x-y$ plane, $ \phi = arctan(y^*/x^*)$, in order to completely describe the steady state. The steady state magnetization is independent of the $\omega_z$ field (Eq.\eqref{eq:fixed.point.mz0.single}). We see that varying  the $\omega_x$ field, both starting from a trivial ($\kappa/\omega_0>1$) or time-crystal phase ($\kappa/\omega_0<1$), induces the appearance of new pairs of steady states on the $x-y$ plane.

\section{Stability analysis ($d=2$)}
\label{sec.app.stabilityd2}

A linear stability analysis of the fixed points of the model can be performed from an spectral analysis of the Jacobian matrix. The approach is a follows. We recall that the dynamical equations of motion of the system can be written as $d m^\alpha /dt = f_{\alpha} (m^x,m^y,m^z)$, with $\alpha=x,y,z$ and $f_{\alpha}$  a nonlinear function on the variables. We can define the displaced variables $u^\alpha = m^\alpha - \alpha^*$ around the fixed points and perform a series expansion. We obtain that,
\begin{equation}
 \left( \frac{d u^x}{dt}, \frac{d u^y}{dt}, \frac{d u^z}{dt} \right)^T = \hat J 
 (u^x, u^y, u^z)^T + O((u^\alpha)^2,u^\alpha u^\beta),
\end{equation}
where $\vec{v}^T$ denotes the transposed of the line vector $\vec{v}$, $\hat J$ is a $3 \times 3$ matrix (denoted as Jacobian matrix) with elements $(\hat J)_{\alpha \beta} = \partial f_\alpha/ \partial \beta$
and the correction terms $O((u^\alpha)^2,u^\alpha u^\beta)$ are quadratic on the displaced variables and can be neglected within a linear approximation. The effective dynamical equations of motion around the fixed point are in this way linear differential equations which can be solved by the eigenspectrum of the Jacobian matrix. The spectral properties of the Jacobian thus provide all information of dynamics around the fixed points, at first order level.
Let us analyse in detail the Jacobian eigenvalues for the two different cases of fixed points in the model.

$\bullet$ Ferromagnetic fixed points, $z^* \neq 0$: for the pair of fixed points $z^*_\pm $ of Eq.\eqref{eq:fixed.point.mzneq0.single} we compute the eigenvalues using sympy package from python. We obtain that their eigenvalues are given by,
\begin{equation}
 \lambda_{[z_\pm^*]} = \left\{0,\quad  z^*_{\pm} \pm  2\frac{\sqrt{\omega_zA}}{\kappa^2 - 4\omega_x\omega_z + 4\omega_z^2} 
        \right\},
        \label{eq:jacobian.eig.mzneq0.single}
\end{equation}
where,
\begin{align}
       A &= \left(\kappa^4\omega_x  + \kappa^2\omega_0^2\omega_z + 16\kappa^2\omega_x\omega_z^2  + 4\omega_0^2\omega_z^3+ 16\omega_x^3\omega_z^2 \right. \nonumber \\ 
	&\quad\:\left.  + 48\omega_x\omega_z^4 \right)
	 - \left( \kappa^4\omega_z + \kappa^2\omega_0^2\omega_x + 8\kappa^2\omega_z^3 +8\kappa^2\omega_x^2\omega_z \right.\nonumber \\ 
	&\quad\:\left.+ 4\omega_0^2\omega_x\omega_z^2 + 48\omega_x^2\omega_z^3  + 16\omega_z^5\right).
\end{align}
Apart from the trivial null eigenvalue, we see that the nontrivial eigenvalue has always a real part. As a consequence the possible steady states of the model with $z^* \neq 0$ are always hyperbolic fixed points. 

$\bullet$ Paramagnetic fixed points, $z^*=0$:
 for the case with $z^*=0$ the Jacobian matrix is simplified and one can obtain analytically their eigenvalues, which have the form,
\begin{equation}
\lambda = \left\{ 0,
\pm\sqrt{B} 
\right\},
\label{eq:jacobian.eig.mz0.single}
\end{equation}
where,
\begin{align}
B &= \left(3\kappa\omega_0 y^* + 8\kappa\omega_x x^*y^*  + 2\omega_0\omega_z x^* + 4\omega_x\omega_z (x^*)^2\right)\nonumber \\ 
&\quad- \left(\omega_0^2 + 2\kappa^2 \mathcal{N} + 4\omega_0\omega_x x^* + 4\omega_x^2 (x^*)^2 + 4\omega_x\omega_z (y^*)^2\right).
\end{align}
In this case we see that the eigenvalues (apart from the trivial one) are purely real or imaginary, thus corresponding either to hyperbolic fixed points or centers. \\

\section{Non Hermitian Perturbation Theory}
\label{sec.app.nonhermPT}

The Jacobian of the system in Sec.\eqref{sec.symm.breaking.pert} is written as 
\begin{equation}
\hat J_\delta = \hat J +\delta \hat V,
\label{eq:jacobian.plus.pert}
\end{equation}
 where $\hat J$ is the unperturbed Jacobian and 
\begin{equation}
 \hat V = \begin{pmatrix}
     0& -1 &0 \\
    1 &0  & 0\\
    0& 0 &0 
    \end{pmatrix},
\end{equation}
is the perturbation matrix. We study the effects of such perturbation in the spectral properties of the Jacobian using Perturbation Theory for general matrices \cite{koch2014}, as follows. 
Any matrix has generalized right ($\vec{u}_i$) and left ($\vec{w}_i$) eigenvectors defined as,
\begin{eqnarray}
\hat J \vec{u}_{i} = \lambda_i \vec{u}_{i}, \qquad
 \hat J^\dagger \vec{w}_{i} = \lambda_i^* \vec{w}_{i},
\label{eq:pt.left.right}
\end{eqnarray}
where left and right eigenvectors obey the ortogonality property $\mathrm{Tr}( \vec{u}_i^\dagger \vec{w}_j) \propto \delta _{ij}$, and $\lambda_i$ are the generalized eigenvectors. We assume that the eigenvalues are non-degenerated, and thus we are dealing with non-degenerate Perturbation Theory. Expanding the eigenvalues and eigenvectors in terms of the perturbative term $\delta$, we can write then as,
\begin{equation}
\lambda_i = \sum_{j = 0}^{\infty} \delta^j \lambda_i^{(j)}, \qquad  
\vec{u}_i = \sum_{j = 0}^{\infty} \delta^j \vec{u}_i^{(j)}, \qquad 
\vec{w}_i = \sum_{j = 0}^{\infty} \delta^j \vec{w}_i^{(j)},
\label{eq:pt.corections}
\end{equation}
where the index $j$ denotes the correction order in perturbation theory, \textit{i.e.} $j=0$ corresponds to the eigenvalues and eigenvectors of the unperturbed Jacobian $J$. Combining Eq.\eqref{eq:jacobian.plus.pert} and Eq.\eqref{eq:pt.corections} into Eq.\eqref{eq:pt.left.right}, one can find the recursive expressions for general $j$'th order corrections. In particular, the corrections to the eigenvalues of the Jacobian are given by,
\begin{equation}
 \lambda_i^{(j)} = \mathrm{Tr}\left( \vec{w}_i^{(0)^\dagger} \hat V \vec{u}_i^{(j-1)} \right) -
 \sum_{k=1}^{j-1} \lambda_i^{(k)} \mathrm{Tr}\left( 
 \vec{w}_i^{(0)} \vec{u}_i^{(j-k)} \right).
\end{equation}

Focusing on the first order terms ($j=1$) we thus  find the equations for the perturbative eigenvalue corrections 
\begin{equation}
 \lambda_i^{(1)} = 
\mathrm{Tr}( \vec{w}_i^{(0)^\dagger} \hat V \vec{u}_i^{(0)} )
\end{equation}

\section{$SU(3)$ structure constants}
\label{app.gelmann}
 
 The explicit form of the non-zero $SU(3)$ structure constants $f_{\rm abc}$ and $d_{\rm abc}$ are listed in Tables \eqref{tab:table1} and \eqref{tab:table2}.

\begin{table}{h!}
  \begin{center}
    \caption{Non-zero structure constants $f_{abc}$ of $SU(3)$ }
    \label{tab:table1}
\begin{tabular}{ c | c c c c c c c c c| c }
\hline
  $abc$ & $f_{abc}$  & & & & & & & & $abc$ & $f_{abc}$ \\
\cline{1-2}
\cline{10-11}
$123$ & $1$  & & & & & & & & $345$ & $\frac{1}{2}$ \\
$147$ & $\frac{1}{2}$  & & & & & & & & $367$ & $-\frac{1}{2}$ \\
$156$ & $-\frac{1}{2}$  & & & & & & & & $458$ & $\frac{1}{2}\sqrt{3}$ \\
$246$ & $\frac{1}{2}$  & & & & & & & & $678$ & $\frac{1}{2}\sqrt{3}$ \\
$257$ & $\frac{1}{2}$  & & & & & & & & & \\
\end{tabular}
\end{center}
\end{table}

\begin{table}[h!]
  \begin{center}
    \caption{Non-zero structure constants $d_{abc}$ of $SU(3)$ }
    \label{tab:table2}
\begin{tabular}{ c | c c c c c c c c c| c }
\hline
  $abc$ & $d_{abc}$  & & & & & & & & $abc$ & $d_{abc}$ \\
\cline{1-2}
\cline{10-11}
$118$ & $\frac{1}{\sqrt{3}}$  & & & & & & & & $355$ & $\frac{1}{2}$ \\
$146$ & $\frac{1}{2}$  & & & & & & & & $366$ & $-\frac{1}{2}$ \\
$157$ & $\frac{1}{2}$  & & & & & & & & $377$ & $-\frac{1}{2}$ \\
$228$ & $\frac{1}{\sqrt{3}}$  & & & & & & & & $448$ & $-\frac{1}{2\sqrt{3}}$ \\
$247$ & $-\frac{1}{2}$  & & & & & & & & $558$ & $-\frac{1}{2\sqrt{3}}$ \\
$256$ & $\frac{1}{2}$  & & & & & & & & $668$ & $-\frac{1}{2\sqrt{3}}$ \\
$338$ & $\frac{1}{\sqrt{3}}$  & & & & & & & & $778$ & $-\frac{1}{2\sqrt{3}}$ \\
$344$ & $\frac{1}{2}$  & & & & & & & & $888$ & $-\frac{1}{\sqrt{3}}$ \\
\end{tabular}
\end{center}
\end{table}


\begin{thebibliography}{99}
\bibitem[{not(A detailed analysis of the initial state dependence is an
  interesting topic. It is, however, beyond the scope of this work)}]{note1}
 (\bibinfo{year}{A detailed analysis of the initial state dependence is an
  interesting topic. It is, however, beyond the scope of this work}).
\end{thebibliography}

\end{document}